\documentclass[apj]{emulateapj}
\usepackage{apjfonts}

\newcommand{\msun}{$M_{\odot}$}

\begin{document}

\def\sarc{$^{\prime\prime}\!\!.$}
\def\arcsec{$^{\prime\prime}$}
\def\arcmin{$^{\prime}$}
\def\degr{$^{\circ}$}
\def\seco{$^{\rm s}\!\!.$}
\def\ls{\lower 2pt \hbox{$\;\scriptscriptstyle \buildrel<\over\sim\;$}}
\def\gs{\lower 2pt \hbox{$\;\scriptscriptstyle \buildrel>\over\sim\;$}}
\def\mbh{$M_{\rm BH}$\ }
\def\mh{$M$\ }
\def\sis{$\sigma$\ }
\def\vvir{$V_{\rm vir}$\ }

\title{On the radiative efficiencies, Eddington ratios, and duty cycles of luminous high-redshift quasars}

\author{Francesco Shankar\altaffilmark{1}, Martin Crocce\altaffilmark{2}, Jordi
Miralda-Escud\'{e}\altaffilmark{3,2,4}, Pablo Fosalba\altaffilmark{2},
and David H. Weinberg\altaffilmark{1}} \altaffiltext{1}{Astronomy
Department, Ohio State University,
    140 W. 18th Ave., Columbus, OH-43210, U.S.A.}
    \altaffiltext{2}{Institut de Ci\`{e}ncies de l'Espai (IEEC-CSIC),
Campus UAB, Bellaterra, Spain}
    \altaffiltext{3}{Instituci\'o Catalana de Recerca i Estudis Avan\c cats,
Barcelona, Spain}
\altaffiltext{4}{Institut de Ci\`{e}ncies del Cosmos, Universitat de Barcelona, Barcelona Spain}

\begin{abstract}
We investigate the characteristic radiative efficiency $\epsilon$,
Eddington ratio $\lambda$, and duty cycle $P_0$ of high-redshift
active galactic nuclei (AGN), drawing on measurements of the AGN
luminosity function at $z=3-6$ and, especially, on recent
measurements of quasar clustering at $z=3-4.5$ from the Sloan
Digital Sky Survey. The free parameters of our models are
$\epsilon$, $\lambda$, and the normalization, scatter, and redshift
evolution of the relation between black hole mass \mbh and halo
virial velocity $V_{\rm vir}$.  We compute the luminosity function
from the implied growth of the black hole mass function and the
quasar correlation length from the bias of the host halos. We test
our adopted formulae for the halo mass function and halo bias
against measurements from the large N-body simulation developed by
the MICE collaboration. The strong clustering of AGNs observed at
$z=3$ and, especially, at $z=4$ implies that massive black holes
reside in rare, massive dark matter halos. Reproducing the observed
luminosity function then requires high efficiency $\epsilon$ and/or
low Eddington ratio $\lambda$, with a lower limit (based on
$2\sigma$ agreement with the measured $z=4$ correlation length)
$\epsilon\ga 0.7\lambda/(1+0.7\lambda)$, implying $\epsilon \ga
0.17$ for $\lambda > 0.25$. Successful models predict high duty
cycles, $P_0\sim 0.2, 0.5$, and $0.9$ at $z=3.1, 4.5$ and $6$,
respectively, and they require that the fraction of halo baryons
locked in the central black hole is much larger than the locally
observed value. The rapid drop in the abundance of the massive and
rare host halos at $z>7$ implies a proportionally rapid decline in
the number density of luminous quasars, much stronger than simple
extrapolations of the $z=3-6$ luminosity function would predict. For
example, our most successful model predicts that the highest
redshift quasar in the sky with true bolometric luminosity $L >
10^{47.5}\, {\rm erg\, s^{-1}}$ should be at $z \sim 7.5$, and that
all quasars with higher apparent luminosities would have to be
magnified by lensing.
\end{abstract}



\section{INTRODUCTION}
\label{sec|intro}

The masses of the central black holes in local galaxies are
correlated with the luminosities, stellar and dynamical masses, and
velocity dispersions of the galaxies in which they reside (e.g.,
Magorrian et al. 1998; Ferrarese \& Merritt 2000; Gebhardt et al.
2000; McLure \& Dunlop 2002; Marconi \& Hunt 2003; H\"{a}ring \& Rix
2004; Ferrarese \& Ford 2005; Greene \& Ho 2006; Graham 2007;
Hopkins et al. 2007b; Graham 2008). The
\mbh-\sis relation, together with the observed correlation between
outer circular velocity and central velocity dispersion measured by
several groups (e.g., Ferrarese 2002; Baes et al. 2003; Pizzella et
al. 2005; Buyle et al. 2006), implies a mean correlation between
black hole mass and the mass or virial velocity of the host galaxy's
dark matter halo, although with a possibly large scatter (e.g., Ho
2007a,b). Recent observational studies have attempted to constrain
the evolution of the black holes and their host galaxies, by
measuring the \mbh-\sis relation at $0< z \lesssim 3$, finding only
tentative evidence for larger black holes at fixed velocity
dispersion or stellar mass (e.g., McLure et al. 2006; Peng et al.
2006; Shields et al. 2006; Treu et al. 2007; Shankar et al. 2009b)
with respect to what is observed locally. However, such an evolution
is difficult to detect given the limited sampling and bias effects
involving these measurements (e.g., De Zotti et al. 2006; Lauer et
al. 2007; Ho 2007a). Probing this evolution becomes even more
difficult at $z>2$ because luminous AGNs substantially outshine
their hosts.

Another way to probe the evolution of black holes and their host
galaxies comes from clustering. Since more massive halos exhibit
stronger clustering bias (Kaiser 1984; Mo \& White 1996), the
clustering of quasars provides an indirect diagnostic of the masses
of halos in which they reside (Haehnelt et al. 1998;
Haiman \& Hui 2001; Martini \& Weinberg 2001; Wyithe \& Loeb 2005),
which in turn can provide information on black hole space densities,
on duty cycles and lifetimes, and, indirectly, on the physical
mechanisms of black hole feeding. Measuring the clustering as a
function of redshift and quasar luminosity probes the relation
between AGN luminosity and host halo mass, thus constraining the
distributions of Eddington ratios and radiative efficiencies which
govern the accretion of black holes at different epochs and in
different environments. The strong clustering of quasars at $z>3$
recently measured by Shen et al. (2007; hereafter S07) in the Sloan
Digital Sky Survey (SDSS; York et al. 2000) quasar catalog (Richards
et al. 2002; Schneider et al. 2007) implies that the massive black
holes powering these quasars reside in massive, highly biased halos.

The classical modeling of quasar clustering by Haiman \& Hui (2001)
and Martini \& Weinberg (2001) assumes a mean value for the duty
cycle and derives the relation between quasar luminosity and host
halo mass by monotonically matching their cumulative distribution
functions. White, Martini \& Cohn (2008; hereafter WMC) have applied
this method to the S07 measurements, concluding that the strong
clustering measured at $z\sim 4$ can be understood only if quasar
duty cycles are high and the intrinsic scatter in the
luminosity-halo relation is small. In this paper, we take a further
step by jointly considering the evolution of the black hole-halo
relation and the black hole mass function, as constrained by the
observed AGN luminosity function and clustering. We examine
constraints on the host halos, duty cycles, radiative efficiencies,
and mean Eddington ratios of massive black holes at $z>3$, imposed
by the clustering measurements of S07 and by a variety of
measurements of the quasar luminosity function at $3 \le z \le 6$
(e.g., Kennefick et al. 1994; Pei 1995; Fan et al. 2001, 2004;
Barger et al. 2003; Wolf et al. 2003; Hunt et al. 2004; Barger \&
Cowie 2005; La Franca et al. 2005; Nandra et al. 2005; Cool et al.
2006; Richards et al. 2006a; Bongiorno et al. 2007; Fontanot et al.
2007; Shankar \& Mathur 2007; Silverman et al. 2008; Shankar et al. 2010a,b).

  Our method of incorporating luminosity function constraints is simple.
We assume the existence of a relation between black hole mass \mbh and halo virial velocity \vvir at high redshift and assume that the slope of this relation is the same as observed locally, but leave its normalization, redshift evolution and scatter as adjustable parameters.
Since the halo mass function is predicted from theory at
every redshift, the evolution of the black hole mass function
follows once the \mbh-\vvir relation is specified. This growth of
black holes is then used to predict the AGN luminosity function, in
terms of the assumed radiative efficiency
$\epsilon/(1-\epsilon)=L/\dot{M}_{\rm BH}c^2$ and Eddington ratio
$\lambda=L/L_{\rm Edd}$ of black hole accretion, which can be
compared to observations. This method inverts the ``continuity
equation'' approach to quasar modeling, in which one uses the
observed luminosity function to compute the implied growth of the
black hole mass function (e.g., Cavaliere et al. 1971; Small \&
Blandford 1992; Yu \& Tremaine 2002; Steed \& Weinberg 2003; Marconi
et al. 2004; Shankar et al. 2004; Yu \& Lu 2004; Shankar, Weinberg,
\& Miralda-Escud\'{e} 2009a, hereafter SWM; Shankar 2009).

We make no specific hypothesis about the mechanisms that trigger
high-redshift quasar activity. Our model simply assumes that a
relation between \mbh and \vvir exists and that it is maintained by
mass accretion that produces luminous quasar activity, assuming no significant
time delay between the two. As detailed below,
simultaneously matching the observed luminosity function and the S07
clustering measurements, especially their $z=4$ correlation length,
is in general quite difficult. Moderately
successful models must share the common
requisites of having low intrinsic scatter in the \mbh - \vvir relation and a
high value of the ratio $\epsilon/\lambda$. 
Although these findings are affected by the model adopted to compute
the halo bias factor, we will show that they do not otherwise depend on the details of
our modeling and can be understood in simple, general terms. 

The mass function and clustering bias of rare, massive halos at high
redshift are crucial inputs for our modeling. We therefore test
existing analytic formulae for these quantities against measurements
from the large N-body simulation developed by the MICE
collaboration, which uses $10^9$ particles to model a comoving
volume of $768\, h^{-1}\, {\rm Mpc}$ on a side. 

  Throughout this paper the following cosmological parameters have
been used, consistent with the best-fit model to WMAP5 data (Spergel
et al. 2007): $\Omega_m=0.25$, $\Omega_L=0.75$, $\sigma_8=0.8$,
$n=0.95$, $h\equiv H_0/100\, {\rm km\, s^{-1}\, Mpc^{-1}}=0.7$,
$\Omega_b=0.044$.

\section{MODEL}
\label{sec|method}

\subsection{AGN BIAS AND LUMINOSITY FUNCTION}
\label{subsec|AGN}

In the local universe, the masses of black holes are tightly
correlated with the velocity dispersion \sis of their parent bulges
(e.g., Ferrarese \& Merritt 2000; Gebhardt et al. 2000; Tremaine et
al. 2002; Shankar \& Ferrarese 2009). This relation has been
recently re-calibrated by Tundo et al. (2007) as
\begin{equation}
\log \left(\frac{\bar{M}_{\rm BH}}{M_{\odot}}\right)=8.21+3.83\log
\left(\frac{\sigma}{200\, {\rm km\, s^{-1}}}\right)\, .
    \label{eq|MbhSigma}
\end{equation}
where we denote the average black hole mass at a fixed $\sigma$
as $\bar{M}_{\rm BH}$. The bulge velocity dispersions are in turn correlated with large
scale circular velocities (Ferrarese 2002; see also Baes et al. 2003
and Pizzella et al. 2005):
\begin{equation}
\log V_c=(0.84\pm 0.09)\log \sigma+(0.55\pm 0.19)\, ,
    \label{eq|sigVvir}
\end{equation}
with \sis and $V_c$ measured in ${\rm km\, s^{-1}}$. For a flat
rotation curve, the disk circular velocity is equal to the halo
virial velocity \vvir. Departures from isothermal halo profiles,
gravity of the stellar component, and adiabatic contraction of the
inner halo can alter the ratio \vvir$/V_c$, but the two quantities
should remain well correlated nonetheless (e. g., Mo et al. 1998; Mo
\& Mao 2004 and references therein). Thus, the
correlations~(\ref{eq|MbhSigma}) and (\ref{eq|sigVvir}) imply a
correlation between black hole mass and halo virial velocity,
although we should expect the \mbh-\vvir relation to have a larger
scatter than the observed \mbh-\sis relation (e.g., Ho 2007b).

As mentioned in \S~\ref{sec|intro}, the models we shall construct
assume that black holes at $z>3$ lie on an \mbh-\vvir relation of
similar form. We parameterize this relation as
\begin{equation}
\bar{M}_{\rm BH}=\alpha \left(\frac{V_{\rm vir}}{300\, {\rm km\,
s^{-1}}}\right)^{4.56}\left(\frac{1+z}{4.1}\right)^{\gamma}\times
1.5\times 10^{8}\, M_{\odot}\, ,
    \label{eq|MbhVvirRelation}
\end{equation}
which corresponds to equations~(\ref{eq|MbhSigma}) and
(\ref{eq|sigVvir}) with \vvir replacing $V_c$. We define $\alpha$ as
the normalization of the \mbh-\vvir relation at $z=3.1$, which
corresponds to the mean redshift in the lower subsample of S07. The
factor $\alpha$ allows both for an offset between the $z=3.1$ and
$z=0$ relations and for a ratio \vvir$/V_c\neq 1$ at $z=0$. For
example, typical disk galaxy models (e.g., Mo et al. 1998; Seljak
2002; Dutton et al. 2007; Gnedin et al. 2007) have
$V_c/$\vvir$\approx 1.4-1.8$ at $z=0$, which would imply
normalizations $\alpha \approx 5-15$ for
equation~(\ref{eq|MbhVvirRelation}) at $z=0$ because of the steep
power of velocity. Note that none of our results depend on
the $z=0$ normalization of the $\bar{M}_{\rm BH}-\sigma$ relation
because we use only high redshift data in this paper. In
addition, we allow redshift evolution in the \mbh-\vvir relation at
$z>3.1$ through the index $\gamma$.

The relation between the halo virial velocity \vvir and the halo
virial mass $M$ is
\begin{eqnarray}
V_{\rm vir}=\left(\frac{GM}{R_{\rm
vir}}\right)^{1/2}=228\left(\frac{M}{10^{12}\, {\rm
M_{\odot}}}\right)^{1/3}\times \nonumber\\
\left[\frac{\Omega_m}{0.25}\frac{1}{\Omega_m^z}\frac{\Delta}{18\pi^2}\right]^{1/6}\left(\frac{1+z}{4.1}\right)^{1/2}\,
{\rm km\, s^{-1}}\, , \label{eq|Vvir}
\end{eqnarray}
where the mean density contrast (relative to critical) within the
virial radius $R_{\rm vir}$ is $\Delta=18\pi^2+82d-39d^2$, with
$d\equiv \Omega_m(z)-1$, and
$\Omega_m(z)=\Omega_m(1+z)^3/\left[\Omega_m(1+z)^3+\Omega_{\Lambda}\right]$
(Bryan and Norman 1998; Barkana \& Loeb 2001). In terms of halo
mass, equation~(\ref{eq|MbhVvirRelation}) corresponds to
\begin{eqnarray}
\bar{M}_{\rm BH}=\alpha \left(\frac{M}{10^{12}\, {\rm
M_{\odot}}}\right)^{1.52}\left[\frac{\Omega_m}{0.25}\frac{1}{\Omega_m^z}\frac{\Delta}{18\pi^2}\right]^{0.76}\times\nonumber\\
\left(\frac{1+z}{4.1}\right)^{\gamma+2.28}\times 4.3\times 10^{7}\,
M_{\odot}\, .
    \label{eq|MbhMhalo}
\end{eqnarray}
We assume the presence of a scatter about this mean relation, with a
log-normal distribution and a dispersion $\Sigma$ in the logarithm
of black hole mass at fixed \vvir.

Given the theoretically known halo mass function, we
compute the black hole mass function via the convolution
\begin{eqnarray}
\Phi_{\rm BH}(M_{\rm BH},z)=\int
\Phi_h(M,z)(2\pi\Sigma^2)^{-1/2}\times \nonumber\\
\exp \left[-\frac{(\log \bar{M}_{\rm BH}[M,z]-\log M_{\rm BH})^2}
{2\Sigma^2}\right]d\log M \, ,
    \label{eq|FMBH}
\end{eqnarray}
with
\begin{equation}
\Phi_s(x,z)=n_s(x,z)x \ln(10)\, ,
    \label{eq|PhibhLog}
\end{equation}
where $\Sigma$ is the log-normal scatter in \mbh at fixed halo mass,
$x=$\mbh$\,$ or $M$, and $n_s(x,z)dx$ is the comoving number
density of black holes/halos (for subscript $s={\rm BH}$ or $s=h$)
in the mass range $x\rightarrow x+dx$, in units of ${\rm Mpc^{-3}}$
for $H_0=70\, {\rm km\, s^{-1}\, Mpc^{-1}}$. The units of $\Phi_s$
are comoving ${\rm Mpc^{-3}}$ per decade of mass. We convert to
these units in order to compare with the data on the AGN luminosity
function.

  The quasar luminosity function $\Phi(L,z)$, expressed in
  the same units as $\Phi_s(x,z)$,
  is modeled according to a simple
prescription where black holes can be in only two possible states:
active or inactive. All black holes that are active accrete with a
single value of the radiative efficiency, $\epsilon$, and of the
Eddington ratio, $\lambda = L/L_{\rm Edd}$, where $L$ is the
bolometric luminosity and
\begin{equation}
L_{\rm Edd}=1.26\times 10^{38}\, {\rm erg\, s^{-1}}
\left(\frac{M_{\rm BH}}{{\rm M_{\odot}}}\right) =
l\left(\frac{M_{\rm BH}}{{\rm M_{\odot}}}\right) ~,
    \label{eq|Lbol}
\end{equation}
is the Eddington luminosity (Eddington 1922). The growth rate of an
active black hole of mass \mbh is $\dot{M}_{\rm BH}=$\mbh$/t_{\rm
ef}$, where the \emph{e}-folding time is (Salpeter 1964)
\begin{equation}
t_{\rm ef}=4\times 10^8\left(\frac{f}{\lambda}\right)\, {\rm yr} \,
,
    \label{eq|tefold}
\end{equation}
where $f=\epsilon/(1-\epsilon)$, and the radiative efficiency is
$\epsilon=L(1-\epsilon)/[\dot{M}_{\rm BH}c^2]$. (Radiative
efficiency $\epsilon$ is conventionally defined with respect to the
mass inflow rate $\dot{M}$, and the black hole mass growth rate
$\dot{M}_{\rm BH}$ is smaller by a factor $1-\epsilon$ because of
radiative losses).

Once the parameters $\alpha$ and $\gamma$ of the \mbh-\vvir relation
are specified, the growth of $\Phi_{\rm BH}(M_{\rm BH},z)$ is
determined by the (theoretically calculable) evolution of the halo
mass function $n_h(M,z)$. We compute the AGN luminosity function
assuming that this growth is produced by accretion with radiative
efficiency $\epsilon$ and Eddington ratio $\lambda$. This method
inverts a long-standing approach to modeling AGN and black hole
evolution in which one calculates the growth of the black hole mass
function implied by the observed luminosity function using a
``continuity equation''$\,$,
\begin{equation}
\frac{\partial n_{\rm BH}(M_{\rm BH},t)}{\partial
t}=-\frac{1}{t_{\rm ef} \ln(10)^2 M_{\rm BH}}\frac{\partial
\Phi(L,z)}{\partial \log L}\,
    \label{eq|continEq}
\end{equation}
(see, e.g., Cavaliere et al. 1971; Small \& Blandford 1992; Yu \&
Tremaine 2002; Marconi et al. 2004; SWM).
Here we ignore the impact of black hole mergers in the evolution of the black hole mass function, because the black hole mass growth via mergers is relatively small, as we show in detail in the Appendix.

Knowing $\Phi_{\rm BH}(M_{\rm BH},z)$, we can invert
equation~(\ref{eq|continEq}) to obtain the luminosity function
\begin{eqnarray}
\Phi(L,z)=\, -\, \ln(10)t_{\rm ef}\int_{\log M_{\rm BH}}^{\infty}
\frac{\partial \Phi_{\rm BH}(M'_{\rm BH},z)}{\partial
z}\frac{dz}{dt}\times \nonumber\\
\left|\frac{d\log L}{d\log M'_{\rm BH}}\right|d\log M'_{\rm BH}\, .
    \label{eq|continEqInv}
\end{eqnarray}
In practice, we integrate equation~(\ref{eq|continEqInv}) up to
black hole masses of $\log M_{\rm BH}/M_{\odot}=11$.
Equation~(\ref{eq|continEqInv}) assumes a strictly monotonic,
scatter-free relation between AGN luminosity and black hole mass.
Therefore in our models the only source of scatter between AGN
luminosity and halo mass is the scatter in the \mbh-\vvir relation.
However, provided that the $L$-\vvir scatter is fairly small (as we
find it must be to explain the observed clustering), we expect that
it makes little difference whether it arises from scatter in
\mbh$\,$ or scatter in $\lambda$.

The average growth rate of all black holes
(active and inactive) of mass \mbh is
$\langle \dot{M}_{\rm BH}\rangle=P_0 M_{\rm BH}/t_{\rm ef}$,
where the duty cycle $P_0$ is the probability that a black hole is
in the active state. In models with a single value of $\lambda$, the
duty cycle is simply the ratio of the luminosity function to the
mass function,
\begin{equation}
P_0(M_{\rm BH},z)=\frac{\Phi(L,z)}{\Phi_{\rm BH}(M_{\rm BH},z)}\,
    \label{eq|P0general}
\end{equation}
where $L=\lambda \, l\, \frac{M_{\rm BH}}{{\rm M_{\odot}}}$. A physically consistent model must have $P_0\le 1$ for all \mbh and $z$, and can be directly computed from
equations~\ref{eq|FMBH} and \ref{eq|continEqInv}.

In addition to the AGN luminosity function, we test our models
against the clustering measurements of S07, specifically their
reported values of the AGN correlation length $r_0$. We calculate
these correlation lengths from the condition
\begin{equation}
\bar{b}^2(z)D^2(z)\xi(r_0)=1\, ,
    \label{eq|r0}
\end{equation}
where $\xi(r_0)$ is the Fourier transform of the linear power
spectrum, $D(z)$ is the linear growth factor of perturbations, and
$\bar{b}(z)$ is the mean clustering bias of AGN shining above a
luminosity threshold $L_{\rm min}$ at redshift $z$, given by (Haiman
\& Hui 2001)
\begin{equation}
\bar{b}(z)=\frac{\int_{L_{\rm min(z)}}^{\infty}\Phi(L,z)b(L,z)d\log
L}{\int_{L_{\rm min}(z)}^{\infty}\Phi(L,z)d\log L}\, .
    \label{eq|bias}
\end{equation}
The minimum luminosity $L_{\rm min}(z)$ in equation~(\ref{eq|bias})
is a bolometric quantity, while the S07 bias is measured above a
redshift dependent, $K$-corrected $M_i$ magnitude. To convert from
magnitudes to bolometric luminosities, we first convert to $B$
magnitudes assuming $M_B=M_i(z=2)+0.804$ (Richards et al. 2006b),
and then adopt an average bolometric correction of $C_B=10.4$, with
$L=C_B L_B\nu_B$, where $\nu_B$ is the frequency at the center of
the $B$-band (at wavelength $4400$ {\AA}). Because our models assume
a single Eddington ratio $\lambda$, $b(L,z)$ is just equal to the
bias $b(M_{\rm BH},z)$ of black holes of mass $M_{\rm
BH}=L/l\lambda$. The latter is computed from the $b(M,z)$ of
\emph{halos} of mass $M$ using the model relation between \mbh and
$M(z)$ (equation~[\ref{eq|MbhMhalo}]). Including the log-normal
scatter of width $\Sigma$, the black hole bias is
\begin{eqnarray}
b(M_{\rm BH},z)=\left[\Phi_{\rm BH}(M_{\rm BH},z)\right]^{-1}\int
b(M,z)\Phi_h(M,z)\times \nonumber\\
(2\pi\Sigma^2)^{-1/2} \exp \left[-\frac{(\log \bar{M}_{\rm
BH}[M,z]-\log M_{\rm BH})^2} {2\Sigma^2}\right]d\log M \, .
    \label{eq|biasL}
\end{eqnarray}
We discuss our choice of $b(M,z)$ in \S~\ref{subsec|comparesimul}.

In summary, the free parameters of our model are:
\begin{itemize}
  \item the normalization constant $\alpha$ in the \mbh-\vvir
  relation,
  \item the parameter $\gamma$ which regulates the redshift evolution
  $[(1+z)/4.1]^{\gamma}$ of this relation,
  \item the mean Eddington ratio $\lambda$ of active black holes,
  \item the log-normal scatter $\Sigma$ in \mbh at fixed \vvir,
  \item the radiative efficiency $\epsilon$ of black hole accretion.
\end{itemize}
The predicted bias in these models is completely independent of the
assumed radiative efficiency (when other parameters are held fixed),
since the efficiency does not affect the relation between luminosity
and halo mass.

\begin{figure}[ht!]
\epsscale{1.1} \plotone{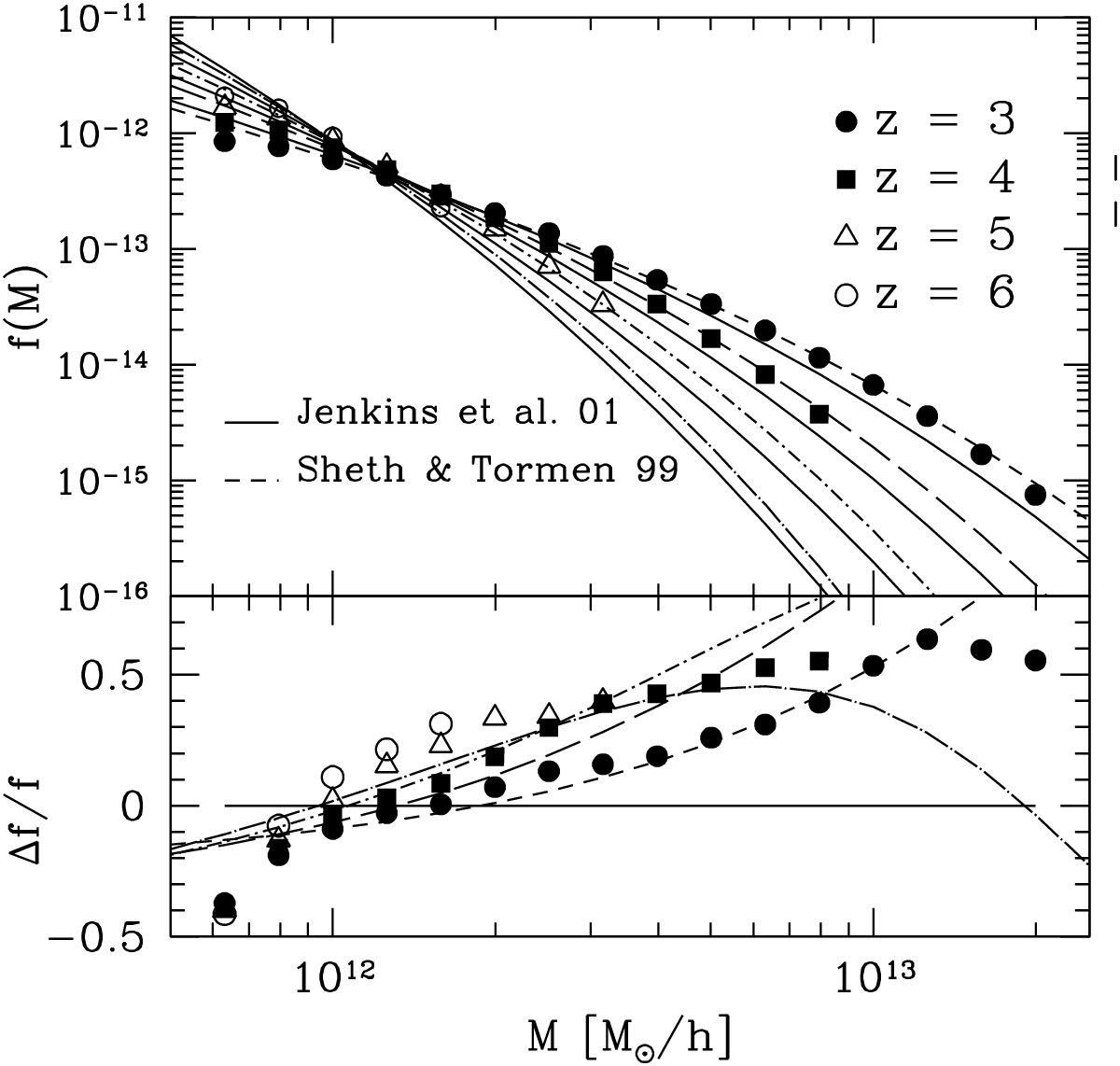} \caption{Halo mass function:
comparison between N-body measurements and analytic fits. {\it Upper
panel :} Measured mass functions for FoF halos identified at
redshifts 3 (\emph{filled circles}), 4 (\emph{filled squares}), 5
(\emph{open triangles}) and 6 (\emph{open circles}) in our
simulation are in better agreement with Sheth \& Tormen (1999)
(\emph{discontinuous} lines) rather than the Jenkins et al. (2001)
(\emph{solid} lines) analytic fits. Fits are displayed for the same
redshifts as measurements in the simulation: from $z=3$
(\emph{upper} lines) to $z=6$ (\emph{bottom} lines). {\it Lower
panel :} Fractional deviations of the measured mass functions
(symbols as in upper panel) and the Sheth \& Tormen predictions
(\emph{discontinuous} lines as above) with respect to the Jenkins et
al. fit (\emph{solid} line).} \label{fig|simulMF}
\end{figure}

\subsection{MASS FUNCTION AND HALO BIAS}
\label{subsec|comparesimul}

The high redshift quasars used by S07 and our study are believed to
reside in very rare halos with ${\rm M}\sim 10^{12-13} h^{-1}\,{\rm
M_{\odot}}$ at redshifts $z=3$ to $6$. While extensive work has been
done to determine the abundances and clustering of halos at $z<3$,
testing the accuracy of simple analytic formulae against predictions
from cosmological numerical simulations of structure formation, this
work has not been extended to the high-redshift ($z\sim 3-6$), rare
massive halos we are interested in here (but see Reed et al. 2007,
2008).

  We perform this test here, using a large N-body simulation from the
MICE collaboration (Fosalba et al.\ 2007) with $1024^3$ particles
and cubic volume of side $L_{\rm box}=768\,h^{-1}\,{\rm Mpc}$, for
the cosmological parameters listed in \S~\ref{sec|intro}. The
initial conditions were set at $z=50$ using the Zel'dovich
approximation, with an input linear power spectrum given by the
analytic fit of Eisenstein and Hu (1999). The subsequent
gravitational evolution was followed using the Tree-SPH code
Gadget-2 (Springel et al. 2005). Halos were identified using the
Friends-of-Friends algorithm (Davis et al.\ 1985), with linking
length equal to $0.164$ times the mean interparticle density.
The minimum halo mass resolved in the simulation is ${\rm M}_{\rm
min}=6\times 10^{11} h^{-1}\, {\rm M_{\odot}}$, with a minimum of 20
particles per halo.

The halo mass function from the simulation is shown as solid symbols
in Figure~\ref{fig|simulMF} (filled circles, filled squares, open
triangles and open circles indicate the abundances of halos at
redshifts $z=$3, 4, 5 and 6, respectively). We plot the quantity
$f(M,z)=n_h(M,z)dM$, where $n_h(M,z)dM$ is the number of halos per
comoving volume at redshift $z$ with mass between $M$ and $M+dM$.
The dashed and solid lines are the analytical models from Sheth \&
Tormen (1999; ST hereafter) and Jenkins et al. (2001; their equation
B2), respectively, plotted at the same redshifts. To better display
the difference between simulations and models, the lower panel of
Figure~\ref{fig|simulMF} shows the fractional deviation with respect
to the Jenkins et al. (2001) fit, with all the lines and symbols as
in the upper panel.

Overall, we find that the ST model fits the simulations in the range
$z=3$ to $5$ within $15\%$ accuracy, while for $z=6$ the error is
about $20\%$ in the range $\log(M/M_{\odot}\,  h^{-1})=12-13$. We
use the ST model in the rest of the paper, because the Jenkins et
al. (2001) formula is clearly a worse fit to the simulation results
in the regime of interest.

As mentioned before, the halos we are interested in are rare, and
studying their clustering properties is therefore difficult. This
``rarity'' can be quantified by means of the peak height $\nu =
\delta_c / \sigma(M,z)$, which characterizes the amplitude of
density fluctuations from which a halo of mass $M$ forms at a given
redshift $z$ (here, $\delta_c=1.686$, and $\sigma(M,z)$ is the
linear overdensity variance in spheres enclosing a mean mass $M$).

 Gao et al.\ (2005) computed the halo bias using the Millenium
simulation (Springel 2005) at redshifts $z=0-5$, but only for halos
collapsing from fluctuations up to $3\sigma$. Angulo et al.\ (2008)
measured the bias of $\sim 4.5\sigma$ halos, but only for $z\le 3$.
In the context of the reionization of the universe, Reed et al
(2008) studied the bias of $< 4\sigma$ halos at redshift $z > 10$.
Additional work on halo bias is presented in Seljak \& Warren
(2004), Cohn \& White (2008), Basilakos et al.\ (2008), and
references therein. In this section we extend these studies to the
regime of our interest, namely $3-5\sigma$ halos at $z=3-6$.

\begin{figure*}[ht!]
\centering \epsscale{1.1} \plotone{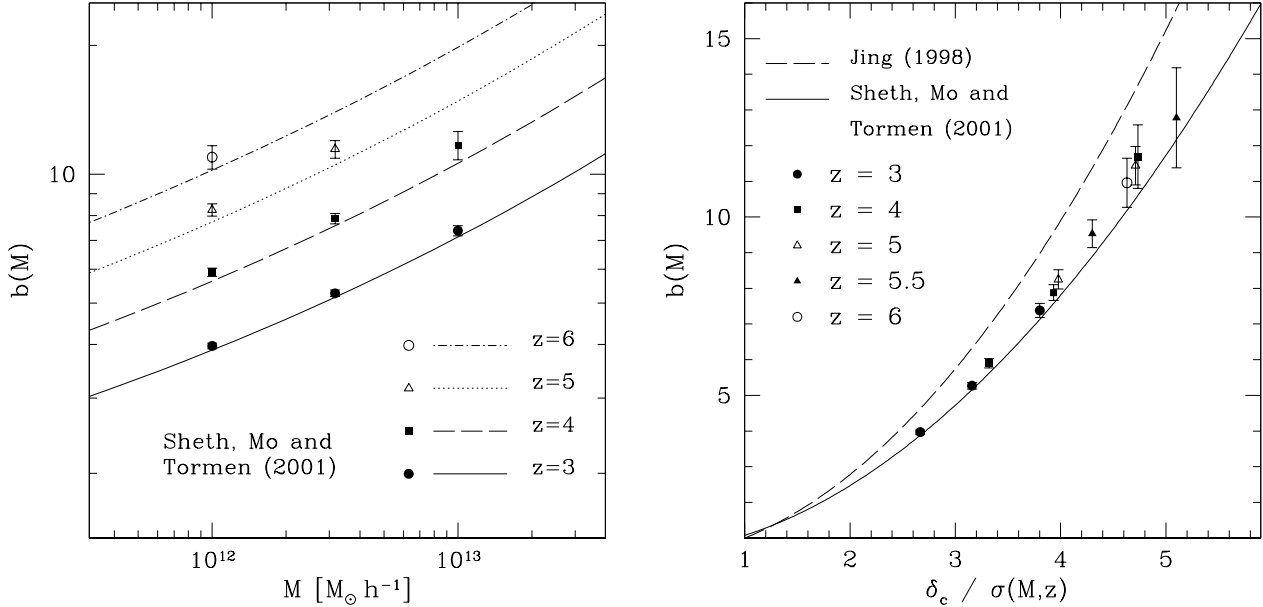} \caption{Halo bias
estimated from $\xi_{hm}/\xi_{mm}$ on scales $8-38\, h^{-1}\, {\rm
Mpc^{-1}}$. The symbols represent the results from a MICE simulation
for different redshifts and halo masses as labeled (see text for
details). In the \emph{left} panel we show halo bias vs. mass and
the corresponding prediction of Sheth, Mo \& Tormen (2001). The
\emph{right} panel shows bias vs. peak height $\nu=\delta_c/\sigma$
and includes the Jing (1998) fit (which mostly coincides with Mo \&
White [1996] expression at these values of $\nu$). The Sheth et al.
(2001) bias works well overall, but it underestimates the results
from the simulation at high redshifts. Jing's fit, on the other
hand, overpredicts the measurements for all masses and redshifts
studied by as much as $15-20\%$.} \label{fig|simul}
\end{figure*}

  We computed the bias factor of halos from simulation outputs at
$z=3,4,5,5.5$ and $6$. At each output we divided the halo catalogue
into three mass bins of equal separation in $\log M$,
$\log(M/M_{\odot} h^{-1})=11.75-12.25, 12.25-12.75$, and
$12.75-13.25$. We then measured the ratio of correlation functions
$b=\xi_{hm}(r)/\xi_{mm}(r)$ at $10$ bins of equal width in $\log r$
in the range $8\,h^{-1}\,{\rm Mpc}\le r \le 38\,h^{-1}\,{\rm Mpc}$
(where $\xi_{hm}$ is the two-point halo-matter correlation and
$\xi_{mm}$ the matter-matter correlation function). The bias was
computed as the mean of these values, and their variance was used as
a rough estimate of the error. We warn that this error indicator may
be underestimating the true uncertainty in our measurements, since
the correlation function errors are correlated in neighboring radial
bins (although this effect is less severe in the presence of shot
noise).

  We analyze halo bias from $\xi_{hm}$ instead of $\sqrt{\xi_{hh}}$ to
overcome the intrinsic noise in the latter quantity due to low halo
abundance (see also Cohn \& White 2008). The two definitions may
differ owing to stochasticity in the halo-matter relation. However,
we have tested that both measures yield consistent results (within
error bars), while the variance among different bins is reduced by
about $50\%$ when using $\xi_{hm}$ (details of this comparison are
given in the Appendix).

In addition to the $8-38\,h^{-1}\,{\rm Mpc}$ measurements, we have
computed bias using the $\xi_{\rm 20}$ measure adopted by S07 and
using the range $30-60\,h^{-1}\,{\rm Mpc}$. These results are
reported in the Appendix. We find no evidence for scale dependence
of the halo bias outside our statistical uncertainties, but the
issue deserves further investigation in future work (Reed et al.
2009).

In Figure~\ref{fig|simul} we show the results for the halo bias at
high redshift as obtained from the MICE simulation (with symbols
corresponding to different redshifts as labeled in the figure). The
left panel depicts the bias as a function of halo mass for various
redshifts; the lines are predictions from the ellipsoidal collapse
formula of Sheth, Mo \& Tormen (2001),
\begin{eqnarray}
b_{SMT}&=&1+\frac{1}{\sqrt{a}\delta_c}\left[\sqrt{a}(a\nu^2)+\sqrt{a} b (a\nu^2)^{1-c} \right. \nonumber \\
&&\left. -\frac{(a\nu^2)^c}{(a\nu^2)^c+b(1-c)(1-c/2)}\right],
\label{SMT01}
\end{eqnarray}
where $a=0.707$, $b=0.5$, $c=0.6$, $\nu=\delta_c/\sigma(M,z)$ and
$\delta_c=1.686$. The right panel shows instead the bias as a
function of peak height $\nu$, in terms of which the predictions for
all redshifts coincide (equation~[\ref{SMT01}]). In addition to
equation~(\ref{SMT01}), we also include in this figure the fitting
formula derived by Jing (1998; see also Mo \& White 1996).

Figure ~\ref{fig|simul} shows that Jing's (1998) fit overestimates
the bias at all redshifts and masses studied at the $15-20\%$ level.
The Sheth, Mo, \& Tormen (2001) prescription is in good agreement
with the simulation for the lower fluctuations ($\nu \leq 4$) that
correspond to halos of mass $3\times 10^{12}\,h^{-1}\,{\rm
M_{\odot}}$ at $z \le 5$, but it underestimates the bias of the
rarest halos of $\nu > 4$ by up to $10\%$. As noted by Crocce,
Pueblas \& Scoccimarro (2006), transients from the Zel'dovich
dynamics generally used to set up the initial conditions lead to
systematically high values for halo bias. This effect manifests
itself more strongly in rare halos, so the discrepancy with Sheth et
al. (2001) in this regime could be a numerical artifact rather than
inaccuracy of the analytic model. Our conclusions are in good
agreement with existing work on halo bias covering slightly
different regimes (and mentioned at the beginning of this section).
Therefore, we will use the Sheth et al. (2001) model for the bias
and discuss in \S~\ref{subsec|bestfitmodel} the impact that adopting
Jing's (1998) bias formula would have on our conclusions.

\begin{figure*}[ht!]
\epsscale{1.1} \plotone{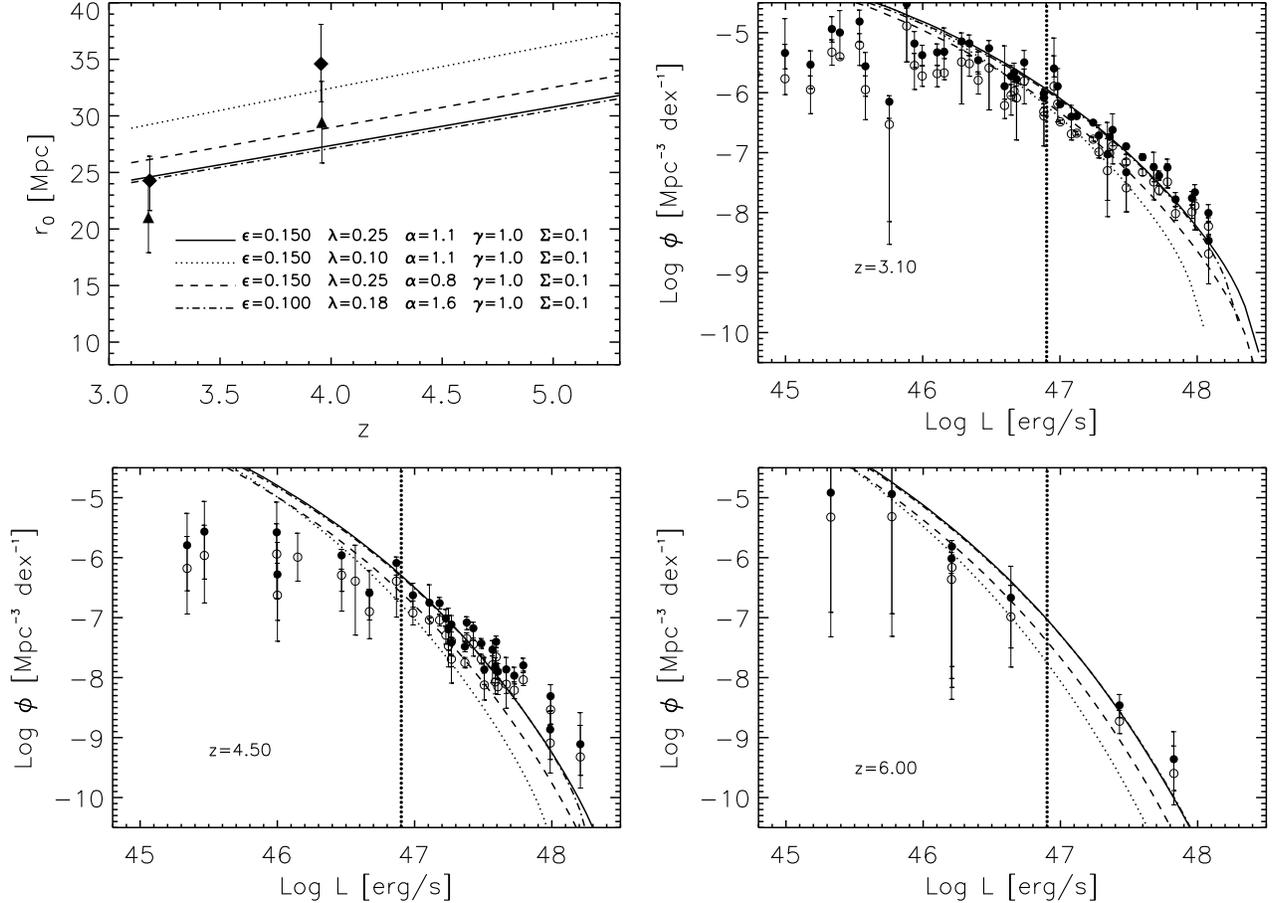} \caption{\emph{Upper left panel}:
Model predictions for the quasar correlation length $r_0$ as a
function of redshift for different values of the input parameters,
as labeled, computed above the luminosity threshold taken from
Figure 1 of Richards et al. (2006a). The \emph{diamonds} and
\emph{triangles} are the Shen et al. (2007) clustering measurements,
corrected to $H_0=70\, {\rm km\, s^{-1}\, Mpc^{-1}}$, calibrated on
their ``good'' and total sample, respectively (see Shen et al. for
details). \emph{Upper right panel}: model predicted luminosity
functions at $z=3.1$, for the same set of models; the data are the
collection from Shankar et al. (2009a) and
Shankar \& Mathur (2007), to which we refer the reader for details.
\emph{Lower left panel}: model predicted luminosity functions at
$z=4.5$. \emph{Lower right panel}: model predicted luminosity
functions at $z=6.0$. The \emph{open} and \emph{filled} circles in
the last three panels represent the AGN luminosity function before
and after obscuration correction, respectively (see text for
details). The vertical, thick, dotted lines in this and the following Figures mark the bolometric luminosity of $L=8\times
10^{46}\, {\rm erg\, s^{-1}}$, taken as the approximate luminosity threshold
of the clustering measurements. Only data in the luminosity function above this threshold have been taken into account in the $\chi^2$-fitting.} \label{fig|r0z}
\end{figure*}

\section{RESULTS}
\label{sec|results}

\subsection{COMPARISON TO OBSERVATIONAL DATA}
\label{subsec|models}

\subsubsection{The data}
\label{subsubsec|data}

In this section we compare our model predictions with the available
data on quasar clustering and the AGN luminosity function. This
comparison is made in Figure 3, where the upper left panel shows
results for the AGN correlation length and the other three panels
show the luminosity function at three different redshifts: $z=3.1$,
$z=4.5$ and $z=6$.

  The data on
the clustering are taken from S07, who have recently extended beyond
$z\sim 3$ previous measurements of the quasar clustering at lower
redshifts from the Two Degree Field Quasar Redshift Survey (Porciani
et al. 2004; Croom et al. 2005; Porciani and Norberg 2006; da
\^{A}ngela et al. 2006; Mounthrichas et al. 2008) and SDSS (e.g.,
Myers et al. 2007; Strand et al. 2008; Padmanabhan et al. 2008). All
the symbols in the upper left panel of Figure~\ref{fig|r0z}
represent the SDSS measurements averaged over sources with redshifts
$2.9\le z \le 3.5$ and $z\ge 3.5$. The diamonds and triangles refer
to the S07 results extracted for the ``good'' and whole samples,
respectively.\footnote{S07 in their clustering analysis remove the
``bad'' fields, i.e. those which do not fully satisfy their
photometric criteria of completeness, but also report clustering
measurements performed on the whole sample. We presume throughout
this paper that the ``good'' measurements (shown with diamonds) are
more reliable and the ones that any successful model must reproduce;
however, following S07, we will always report both sets of data in
the Figures. The smaller number of pairs at small separations in the
``good'' sample could lead to systematic errors in the correlation
length estimate as well as larger statistical uncertainties (Y.
Shen, private communication).} Following S07, we take $z=3.1$ and
$z=4.0$ as the effective measurement redshifts for the two redshift
bins.

The S07 clustering measurements are for optically identified AGNs
only. However, the growth of black holes is connected to the total
luminous output of the AGN population, not that of obscured or
unobscured sources alone. We consider obscuration here as a random
variable not linked with the large scale clustering of AGNs, so that
the correlation length of obscured AGNs is the same as that of
unobscured ones of the same bolometric luminosity. This assumption
is plausible regardless of whether obscuration is principally a
geometrical effect or an evolutionary phase.

Following SWM, we take the AGN luminosity function at the mean
redshifts of $z=3.1$, $4.5$, and $6$, where most of the high
redshift optical and X-ray data sets collected in SWM and Shankar \&
Mathur (2007; and references therein) are concentrated. We then
adopt\footnote{We also insert a Jacobian correction factor in their
equation (4) between observed $B$-band and bolometric luminosities.}
equation~(4) of Hopkins et al. (2007a) to re-normalize the
luminosity function to include obscured sources, assuming the
obscuration is independent of redshift. However, because the
obscuration correction may suffer from significant uncertainties,
even up to a factor of a few (e.g., Ueda et al. 2003; La Franca et
al. 2005; Tozzi et al. 2006; Gilli et al. 2007; Hopkins et al.
2007a), when comparing model predictions to the bolometric
luminosity function we will also include uncertainty in the
obscuration correction as a source of systematic error to be added
to the statistical error of the luminosity function measurements. In
Figure~\ref{fig|r0z} the filled and open circles show the luminosity
function with and without obscuration corrections, respectively.

\subsubsection{General properties of the models}
\label{subsubsec|generalproperties}

\begin{figure*}[ht!]
\epsscale{1.1} \plotone{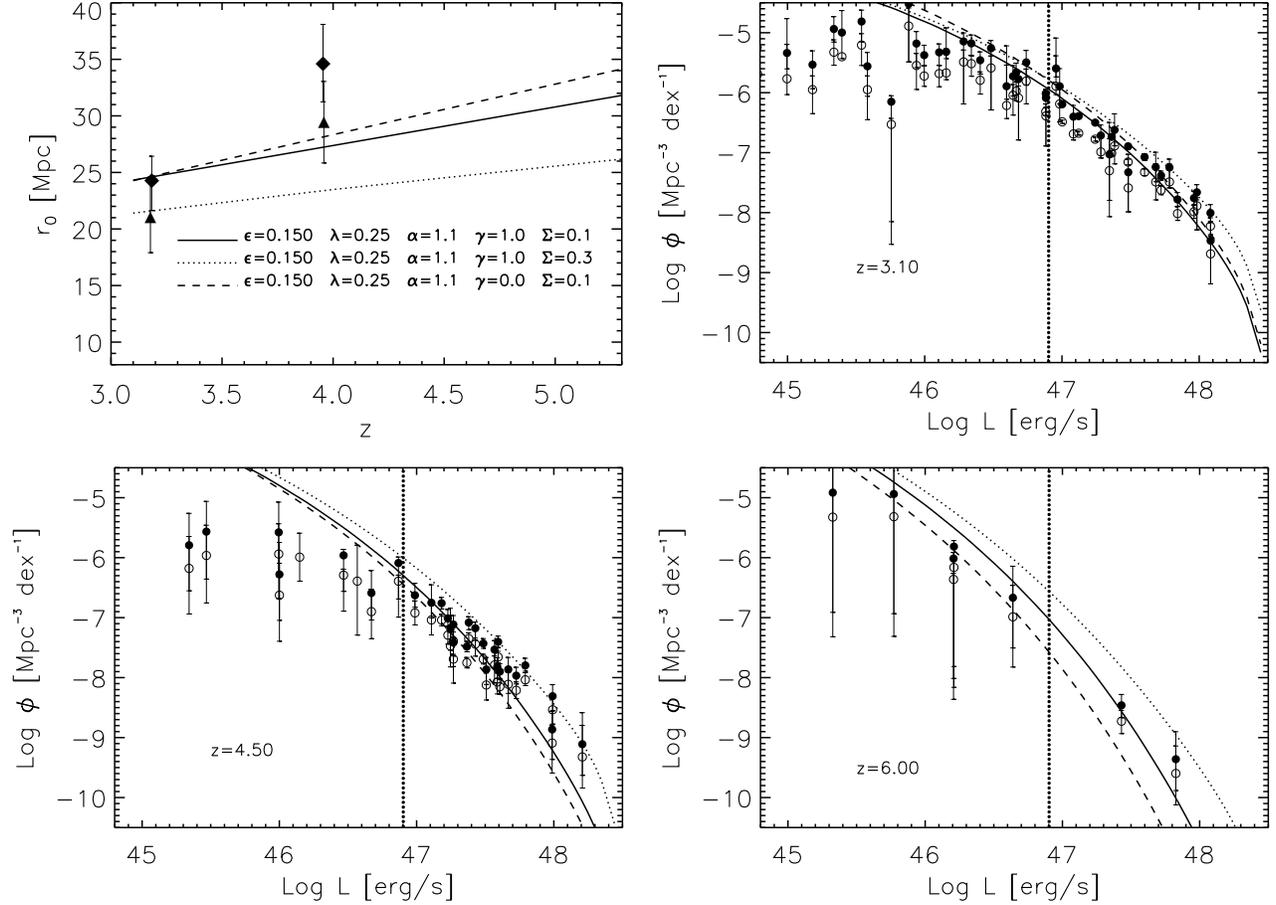} \caption{Impact of changing the
scatter $\Sigma$ or evolution parameter $\gamma$ of the \mbh-\vvir
relation. The format is as in Figure~\ref{fig|r0z}, and the adopted
model parameters are labeled in the upper left panel. Increasing
scatter worsens the match to the clustering data, and lowering
$\gamma$ worsens the match to luminosity function evolution.}
\label{fig|scatter}
\end{figure*}

Figures~\ref{fig|r0z} and \ref{fig|scatter} illustrate the
dependence of model predictions on the adopted parameters. In
general terms, we can understand the interplay between the different
parameters by combining equations~(\ref{eq|MbhMhalo}),
(\ref{eq|tefold}) and (\ref{eq|continEqInv}). Before examining the
impact of individual parameter changes, we should note that there is
one exact degeneracy within our family of models, if the luminosity
function and correlation length are the only constraints. If we
lower the Eddington ratio by a factor of $\Gamma$ but raise the
\mbh-\vvir normalization $\alpha$ by the same factor, then the host
halo mass at a given quasar \emph{luminosity} is unchanged, so the
predicted clustering is unchanged. All black hole masses are larger
by a factor $\Gamma$, and so are their average growth rates required
to match the evolving halo mass function, but if we lower the
efficiency factor $f=\epsilon/(1-\epsilon)$ by the same factor
$\Gamma$, then the luminosity function implied by this growth is
unchanged. Our analysis therefore cannot constrain $\lambda$ and $f$
individually, but it can provide interesting constraints on the
\emph{ratio} $f/\lambda$.

In Figure~\ref{fig|r0z}, all models have evolution parameter
$\gamma=1.0$ and a tight correlation between \mbh and $M$, with
$\Sigma=0.1$ dex. Solid lines in each panel show the predictions of
a ``reference model'' with radiative efficiency $\epsilon=0.15$,
Eddington ratio $\lambda=0.25$, and a normalization of the
\mbh-\vvir relation $\alpha=1.1$. This model matches the S07 value
of $r_0$ at $z=3.1$. At $z=4$, it is consistent with S07's
measurement from the full quasar sample, but it falls below the
``good'' sample measurements by about $2\sigma$. This model is in
fairly good overall agreement with the bright end of the AGN
luminosity function at all redshifts, though it is somewhat low at
$z=4.5$.

In general all our models tend to overpredict the faint end of the
AGN luminosity function below $L\sim 10^{46}\, {\rm erg\, s^{-1}}$
at $z=3.1$ and, more severely, at $z=4.5$. These behaviors suggest
that one or more of the model assumptions break down at lower
luminosities. For example, the assumption of a constant $\lambda$
and $\epsilon$ may not be valid. Alternatively, the assumed
monotonic relation between black hole mass and halo mass could break
down in this regime (see, e.g., Tanaka \& Haiman 2008). However,
these hypotheses cannot be tested with the present data because the
bias measurements by S07 do not probe luminosities fainter than
$L\le 10^{47}\, {\rm erg\, s^{-1}}$, which is where our models start
diverging from the data. We therefore do not attempt to reproduce
the faint end of the AGN luminosity function with our models in this
work.

 We now examine the consequences of varying each of the five model
parameters, as listed at the end of \S~\ref{subsec|AGN}. We first
consider lowering the Eddington ratio to $\lambda=0.1$, keeping the
other parameters fixed. Since the black hole abundances and their
growth rates are fixed by their correspondence to halos, the duty
cycles must increase as the inverse of $\lambda$ to compensate for
the lower accretion rates during the active phase, thereby keeping
the average volume emissivity from quasars constant. The results for
this case are shown as the dotted line in Figure~\ref{fig|r0z}. A
better match to the high observed clustering amplitude is clearly
achieved, because the observed quasars correspond to more massive
black holes and rarer halos. However, the fit to the luminosity
function is worse because the abundance of the most luminous quasars
is underpredicted. Low values of the Eddington ratio are also
disfavored by other observational (e.g., McLure \& Dunlop 2004;
Vestergaard et al. 2004; Bentz et al. 2006; Kollmeier et al. 2006;
Kurk et al. 2007; Netzer \& Trakhtenbrot 2007; Shen et al. 2008) and
theoretical studies (Shankar et al. 2004; Lapi et al. 2006;
Volonteri et al. 2006; Li et al. 2007; SWM; Di Matteo et al. 2008).

 When the \mbh-\vvir normalization $\alpha$ is lowered (dashed curve in
Figure~\ref{fig|r0z}), the effect is simply to lower the black hole
masses and quasar luminosities at fixed abundance. At fixed quasar
luminosity, the clustering increases owing to the greater mass of
the associated halos, but the decrease in abundance prevents a good
match to the data. The dot-dashed curve illustrates the triple
degeneracy described at the beginning of this section: a different
set of ($\lambda$,$\alpha$,$\epsilon$) values whose predictions are
nearly identical to those of the reference model.

Figure~\ref{fig|scatter} shows the effect of varying the scatter
$\Sigma$ or the evolution parameter $\gamma$ of the \mbh-\vvir
relation. A low scatter maximizes the bias for a given set of other
parameters, so low scatter is favored to reproduce the high values
measured for $r_0$ at these redshifts (see also WMC). On the other
hand, semi-empirical studies and AGN theoretical modeling support a
significant intrinsic scatter for the \mbh-$M$ relation at redshifts
$z\lesssim 3$ (e.g., Lapi et al. 2006; Haiman et al. 2007; Myers et
al. 2007; Gultekin et al. 2009). Increasing the scatter to $\Sigma=0.3$ (dotted lines),
boosts the AGN luminosity function by increasing the number of
massive black holes, but it depresses clustering because more
quasars at a given $L$ reside in less massive halos. It also
slightly flattens the dependence of the predicted $r_0$ on redshift.
We then need to lower $\alpha$ or $\lambda$ to restore the
luminosity function and increase $r_0$. However, lowering $\lambda$
or $\alpha$ would also require higher radiative efficiencies to keep
the match to the luminosity function. We also find that models with
$\Sigma=0.3$, $\epsilon \gtrsim 0.25$, and $\alpha \lesssim 0.7$
predict more AGNs than black holes, yielding the unphysical
condition of $P_0>1$ at $z>4$.

The parameter $\gamma$ regulates the amplitude of the model AGN
luminosity function at high redshifts relative to that at $z=3.1$.
Dashed lines in Figure~\ref{fig|scatter} show a model with evolution
index $\gamma=0$ and other parameters the same as those of the
reference model. Lowering $\gamma$ maps the same $L$ to higher mass,
more biased halos at higher redshifts, steepening the $r_0-z$
relation and bringing it closer to the observed trend. However, more
massive halos are rarer at higher redshifts, so the predicted AGN
luminosity function drops significantly below the data at $z=4.5$
and, especially at $z=6$. 
Reproducing the observed luminosity evolution requires positive
evolution ($\gamma
>0$) of the \mbh-\vvir relation.

\subsubsection{A closer comparison}
\label{subsubsec|chi2}

\begin{figure*}[ht!]
\epsscale{1.1} \plotone{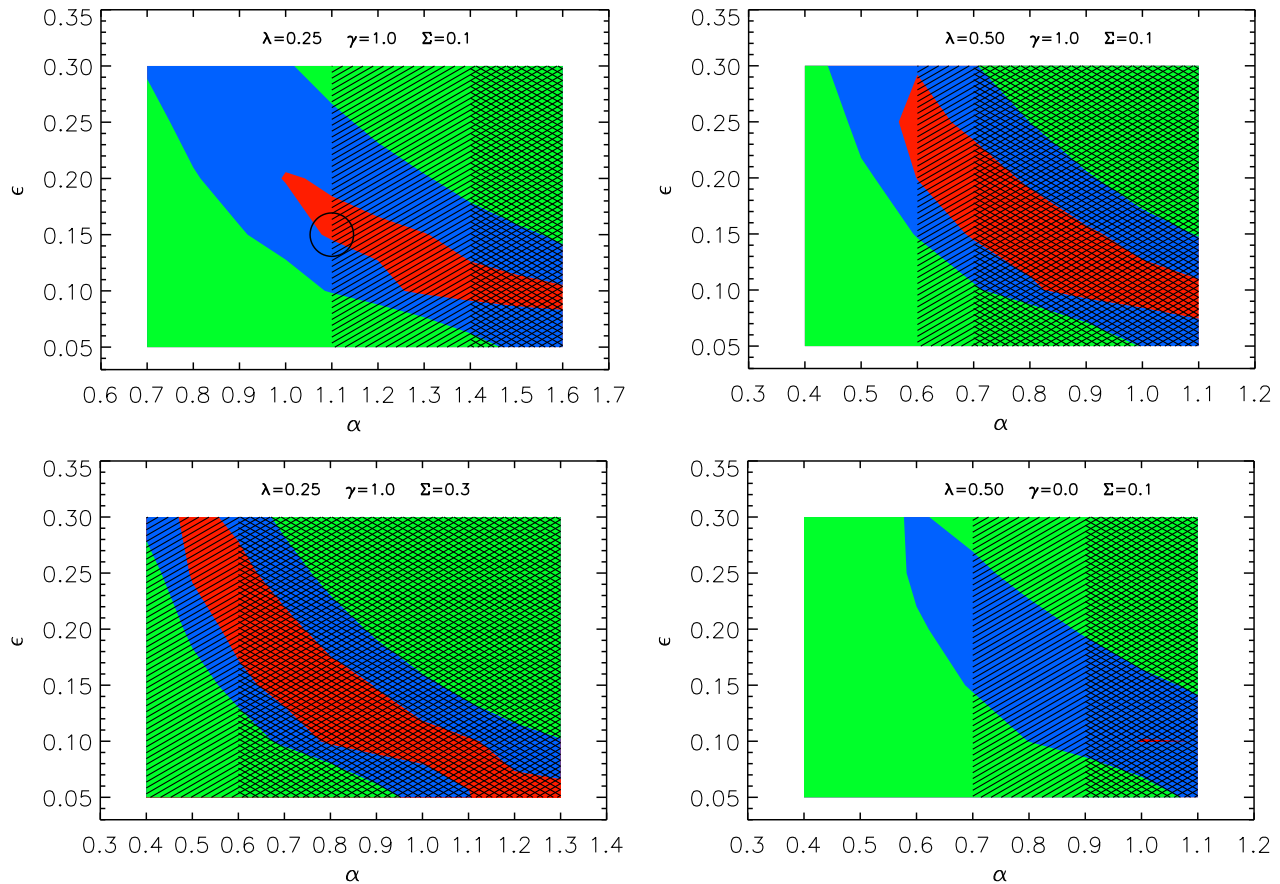} \caption{The $\chi^2_{\rm dof}$
per degree of freedom as a function of the radiative efficiency
$\epsilon$ and  $\alpha$, the normalization of the \mbh-\vvir
relation, with other parameters fixed at the values listed on top of
each panel. The blue and red areas define the regions in the $\epsilon$-$\alpha$ plane
where the $\chi^2_{\rm dof}$ for the luminosity is below 3 and 1.5,
respectively. For the
luminosity function we have used only the data with $L\ge 8\times 10^{46}\,
{\rm erg\, s^{-1}}$, which is the luminosity threshold above which
clustering measurements are available. The double-hatched and hatched
areas define the regions where the $\chi^2_{\rm dof}$
for the correlation length at $z=4$ is above 6 and 4, respectively.
The \emph{circle} in the
upper left panel marks the parameters of our reference model.}
\label{fig|Chi2}
\end{figure*}

Figure~\ref{fig|Chi2} presents a more systematic view of the
dependence of clustering and luminosity function predictions on
model parameters. Because none of our models reproduce all aspects
of the data, and because observational errors may in some cases be
dominated by systematic rather than statistical uncertainties, we
have taken only a semi-quantitative route to comparing models and
measurements. The upper left panel shows models with $\lambda=0.25$,
$\Sigma=0.1$, and $\gamma=1$, defining the contour levels of acceptable
models on a grid of
($\alpha$, $\epsilon$) values.
The blue and red areas define the regions
where the $\chi^2_{\rm dof}$ for the luminosity is below 3 and 1.5,
respectively. Here $\chi^2_{\rm dof}=\chi^2/N$, where $N=45$ is the number of points in
the luminosity function, which include only those points
(from Figure~\ref{fig|r0z}) with $L\ge 8\times
10^{46}\, {\rm erg\, s^{-1}}$, the approximate luminosity threshold
of the clustering measurements, marked with vertical, thick, dotted lines in the Figures. For comparison, note that the models shown by the
dashed and dotted lines in Figure~\ref{fig|r0z} have $\chi^2/N=5.41$
and $18.13$, respectively, while the reference model has
$\chi^2/N=1.37$. If the data points were independent, then even
$\chi^2/N=1.37$ for $N=45$ would be an enormous statistical
discrepancy, but the systematic uncertainty in the obscuration
correction, at least, is highly correlated among points at a given
redshift, motivating our rather loose criterion for ``agreement''.
We assign observational errors to
each data point equal to the reported statistical error (usually
derived from the Poisson error on counts in the bin) summed in
quadrature with 50\% of the difference between the obscuration
corrected and uncorrected luminosity function estimates. This
procedure is \emph{ad hoc}, but it captures the reasonable
expectation that the uncertainty on the obscuration correction is of
the same order as (but smaller than) the correction itself, and the
fact that the scatter among data sets visible in
Figure~\ref{fig|r0z} is comparable to the difference between open
and filled symbols.

The double-hatched and hatched
areas define the regions where $\chi^2=(r_{\rm 0, obs}-r_{\rm 0,
pred})^2/\sigma_{\rm obs}^2$, with the S07 value of $(r_{\rm
0, obs}, \sigma_{\rm obs})=(24.3, 2.4)h^{-1}$ at $z=4.0$, is above 6 (i.e., a $\gtrsim 2.5\sigma$ discrepancy) and 4, respectively.
Note that the contour plots for the clustering are vertical,
given that the predicted clustering strength is independent
of the values for the radiative efficiency (see \S~\ref{sec|method}).
We find the constraints from the S07 clustering measurement of $(r_{\rm 0,
obs}, \sigma_{\rm obs})=(16.9, 1.7)h^{-1}$ at $z=3.1$ not
to be very constraining given that almost all models
explored in Figure~\ref{fig|Chi2} are consistent with
such data at the $\lesssim 2\sigma$ level.
It is clear from Figure~\ref{fig|Chi2} that the acceptable models
are in general places in the upper-left corner of the ($\alpha$, $\epsilon$) plane,
characterized by higher radiative efficiencies and
lower values of $\alpha$ for the same quasar luminosity/black hole mass,
which implies higher halo masses (equation~[\ref{eq|MbhMhalo}])
and corresponding clustering amplitude.

Examination of Figure~\ref{fig|Chi2} reinforces the generality of
the points made in our discussion of Figures~\ref{fig|r0z} and
\ref{fig|scatter}. The circle in the upper left panel marks our
reference model with $\alpha=1.1$, $\gamma=1.0$, $\Sigma=0.1$,
$\lambda=0.25$, $\epsilon=0.15$. Note that our reference model is
\emph{not} the best-fit model, as some models characterized by
higher radiative efficiency and lower $\alpha$ have an overall lower
$\chi^2$. However, we preferred to adopt as working models those
defined by not too extreme values of the radiative efficiency. Also,
the reference model already predicts $P_0\approx 1$ at $z=6$ (see
Figure~\ref{fig|dutycycle} below), and lowering $\alpha$ reduces the
black hole space density and pushes $P_0$ above unity. Other models
with the same $\alpha$ but different $\epsilon$ have
identical clustering, but the match to the observed luminosity
function becomes worse for $\epsilon \le 0.1$ and $\epsilon \ge 0.25$.
Lowering $\alpha$ at fixed $\epsilon$ improves the clustering
agreement but quickly makes the luminosity function agreement worse.
Raising $\alpha$ to 1.2 or 1.3 slightly improves the luminosity
function agreement but worsens the clustering agreement. Raising
$\lambda$ to 0.5 (upper right panel) worsens the agreement with the
$z=4$ clustering if $\epsilon$ and $\alpha$ are held fixed. However,
because of the 3-way degeneracy noted at the beginning of this
section, a model with $\lambda=0.5$, $\epsilon=0.25$, and
$\alpha=0.5$ makes very similar predictions to a model with
$\lambda=0.25$, $\epsilon=0.15$, $\alpha=1.0$ (which has $f/\lambda$
smaller by a factor of $\approx 2$), and we disfavor the higher
$\lambda$ models only on physical grounds because of the high
required efficiency. Models with $\Sigma=0.3$ (lower left) yield
consistently worse agreement with the $z=4$ correlation length
unless lower values of $\alpha$ are adopted, but low-$\alpha$ models
produce $P_0>1$ at high redshifts and require high radiative
efficiencies to match the luminosity function. Models with
$\gamma=0.0$ (lower right) yield consistently worse agreement with
the luminosity function.

Our reference model underpredicts the S07 $z=4$ correlation length
(the value for ``good'' fields) by $2.2\sigma$, and it slightly
underpredicts the observed luminosity function in the luminosity
range corresponding to the S07 quasar sample. If we take these
discrepancies as a maximal allowed level of disagreement, then our
reference model effectively defines a lower limit on the allowed
value of $f/\lambda$, at $f/\lambda$=0.7. This conclusion does not
depend on our adopted bolometric correction. If we assumed a
bolometric correction higher by a factor $\Gamma$, then we would
require higher $\lambda$ (by the same factor) for fixed black hole
masses to match our revised estimate of the bolometric luminosity
function. We would also require higher $f$, again by a factor
$\Gamma$, to reproduce the observed luminosity function history
while building the same black hole population. For observationally
estimated Eddington ratios $\lambda\gtrsim 0.25$ (Kollmeier et al.
2006; Shen et al. 2008) our limit on $f/\lambda$ implies
$\epsilon\gtrsim 0.15$, significantly higher than the radiative
efficiency $\epsilon\approx 0.1$ expected for the disk accretion
onto a non-rotating black hole.

\subsubsection{Varying the bias}
\label{subsubsec|varyingbias}

\begin{figure*}[ht!]
\epsscale{1.1} \plotone{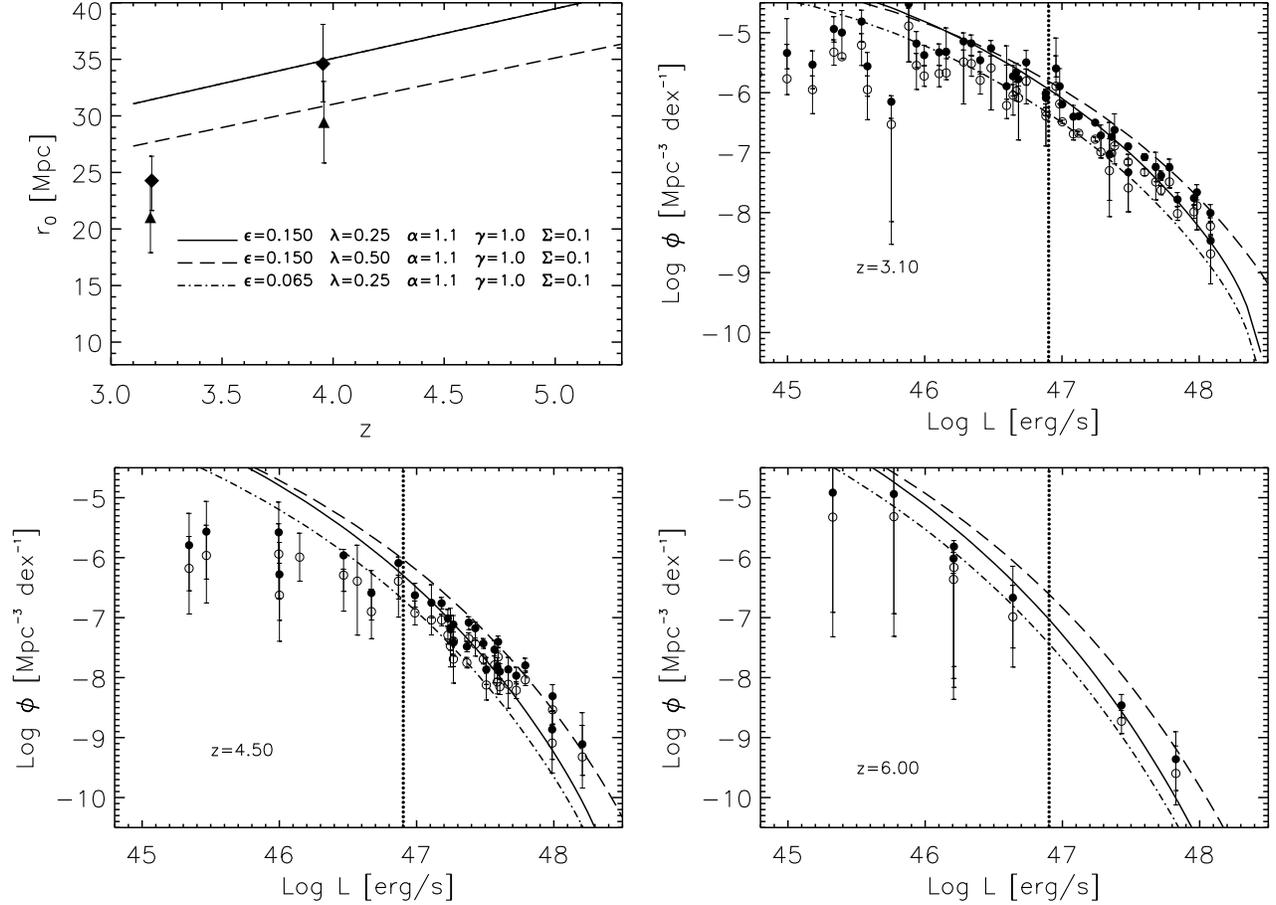}
\caption{Same format as Figure~\ref{fig|r0z}, but Jing's (1998) bias
formula has been adopted instead of the Sheth et al. (2001) one, and
we consider models with lower values of $\epsilon/\lambda$ in
addition to the reference model (\emph{solid} line). In the upper
left panel, the \emph{solid} line overwrites the \emph{dot-dashed}
line because the two models have the same black hole mass-halo mass
relation.} \label{fig|JingModel}
\end{figure*}

These constraints would be much looser if we adopted the Jing (1998)
bias function instead of the Sheth et al. (2001) formula that fits
our N-body data. As already discussed in the previous sections, the
Jing (1998) formula predicts a significantly higher value of the
bias. Therefore, a much larger family of models can match the $z=4$
S07 clustering measurements, with no strict requirement for a high
$f/\lambda$ ratio. For example, Figure~\ref{fig|JingModel} compares
the predictions of the reference model to two alternatives, one with
$\epsilon =0.065$ ($f/\lambda\approx 0.28$) and one with
$\lambda=0.5$ ($f/\lambda\approx 0.22$), with $r_0$ calculated using
the Jing (1998) formula in all models. The luminosity function
predictions of the reference model are unchanged, and all three
models yield acceptable agreement with the $z=4$ clustering
measurement. The low $\epsilon$ model underpredicts the luminosity
function, but the $\lambda=0.5$ model overpredicts it, and lowering
$\epsilon$ to $\sim 0.1$ in this case would yield acceptable
agreement. All three models overpredict the $z=3.1$ correlation
length. With optimal choices of $\alpha$ and $\lambda$, one could
find models that graze the top of the $z=3.1$ error bar and the mean
of the $z=4$ correlation length while acceptably matching the bright
end of the luminosity function (e.g., $\lambda=0.3$,
$\epsilon=0.06$, $\gamma=1$, $\alpha=2.3$, $\Sigma=0.1$).
Alternatively, one could adopt any of the models shown in
Figure~\ref{fig|JingModel} but drive down the $z=3.1$ clustering by
assuming that the scatter $\Sigma$ grows substantially between $z=4$
and $z=3$.

\subsection{BIAS AND DUTY CYCLE PREDICTIONS FOR THE REFERENCE MODEL}
\label{subsec|bestfitmodel}

\begin{figure}[ht!]
\epsscale{1.1} \plotone{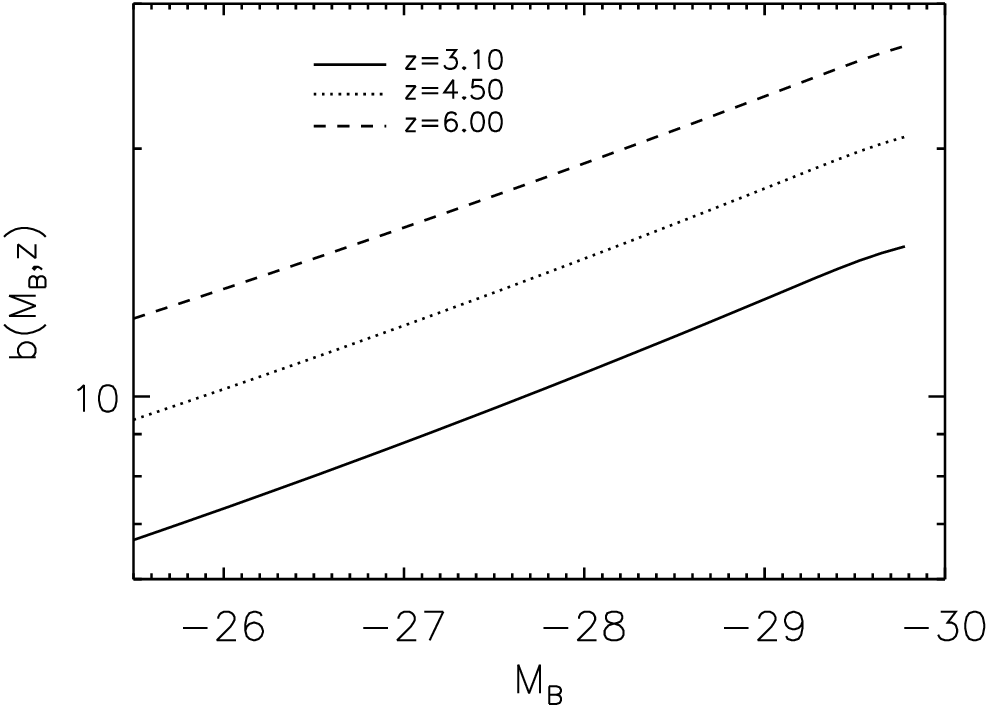}
\caption{Predicted bias for our reference model as a function of
$B$-band magnitude at different redshifts, as labeled.}
\label{fig|bLz}
\end{figure}

Here we discuss further properties and predictions of a model that
simultaneously matches the observed luminosity function and (at the
$2\sigma$ level) the clustering. For simplicity all the results
presented below are obtained from the reference model, which has
($\alpha$, $\Sigma$, $\gamma$, $\lambda$, $\epsilon$)=(1.1, 0.1,
1.0, 0.25, 0.15). Figure~\ref{fig|bLz} shows the predicted bias
$b(L,z)$ as a function of $B$-band magnitude; the solid, dashed and
dotted lines correspond to $z=3.1$, 4.5 and 6, respectively. Note
that $b(L,z)$ now refers to bias at a given luminosity rather than
above a given luminosity. The predicted bias increases significantly with
luminosity, at variance with what has been observed at lower
redshifts $z\lesssim 2$, where evidence for a much flatter behavior
of the bias against luminosity has been found (e.g., da \^{A}ngela
et al. 2006; Myers et al. 2007; Porciani \& Norberg 2006; Coil et
al. 2007; Mountrichas et al. 2008; Padmanabhan et al. 2008). Note
that our prediction applies only to the very bright end of the AGN
luminosity function, with $L \gtrsim 8\times 10^{46}\, {\rm erg\,
s^{-1}}$; there are no available clustering measurements below this
luminosity at $z>3.5$.

We compute the effective halo mass that corresponds to the quasar
hosts at the S07 luminosity thresholds via the relation $b(M_{\rm
eff},z)=\bar{b}(z)$. Our reference model yields $M_{\rm eff}\sim
1.1\times 10^{13}\, {\rm h^{-1}\, M_{\odot}}$, nearly constant
within $3\le z\le 6$. This mass scale is in marginal agreement with
what has been inferred from the clustering analysis of large
AGN-galaxy surveys at lower redshifts (e.g., Myers et al. 2007;
Mountrichas et al. 2008), supporting a roughly constant host halo
mass for luminous quasars at all times. Our reference model
underpredicts the S07 correlation length at $z=4$, so if we used
their measured bias we would obtain a somewhat higher $M_{\rm eff}$
at this redshift. In fact, S07 find an effective host halo mass for
quasars at $z\gtrsim 3.5$ that is a factor of two higher than the
host halo mass for quasars with $2.9\le z \le 3.5$. Their mean
values are $M_{\rm eff}=2.5\times 10^{12}\,h^{-1}\, {\rm M_{\odot}}$
and $5\times 10^{12}\,h^{-1}\, {\rm M_{\odot}}$ in the low and high
redshift bins, respectively, significantly lower than our quoted
value, owing to their use of the Jing (1998) bias formula, which
yields lower halo masses at fixed bias.

Francke et al. (2008) have recently measured the clustering of 58
X-ray selected AGNs at $z\sim 3$ in the Extended Chandra Deep Field
South, cross correlating them with a sample of 1385 luminous blue
galaxies at the similar redshifts. Their quoted bias is $b=4.7\pm
1.7$ corresponding to halos of mass $\log
M/M_{\odot}=12.6^{+0.5}_{-0.8}$, in line with the previous findings
by Adelberger \& Steidel (2005) derived from optical quasar samples
at similar redshifts and luminosities. These studies probe AGNs about 6
magnitudes fainter than those probed by S07, and they seem to
support a significant decrease of the bias at lower luminosities.
These results would then be in qualitative agreement with our model
predictions, but at variance with the flat dependence of quasar
clustering on luminosity found at lower redshifts (e.g., Myers et
al. 2007; Coil et al. 2007; Padmanabhan et al. 2008). Larger surveys
of low-luminosity quasars are needed to reduce the errors on the
clustering measurements.

\begin{figure}[ht!]
\epsscale{1.1} \plotone{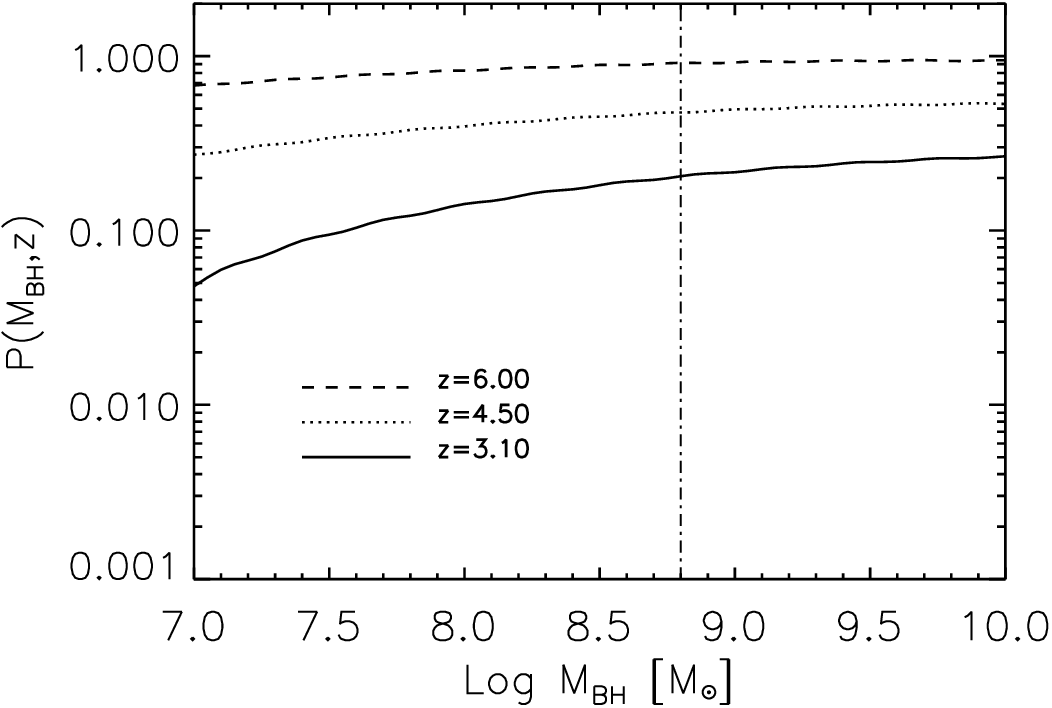}
\caption{Predicted duty cycle for our reference model as a function
of black hole mass at different redshifts, as labeled. The vertical
\emph{dot-dashed} line marks the point below which the results
should be treated with caution, since there are no clustering
constraints below this limit and the model overpredicts the
luminosity function.} \label{fig|dutycycle}
\end{figure}

Figure~\ref{fig|dutycycle} shows the duty cycle $P_0$ as a function
of black hole mass and redshift (eq.~\ref{eq|P0general}). Results
below $\sim 6\times 10^8\, {\rm M_{\odot}}$, marked with a vertical
dot-dashed line in the Figure, should be treated with caution, since
there are no clustering constraints in this regime and the model
overpredicts the luminosity function. Above this limit the predicted
duty cycles are roughly constant with mass, with values of $P_0\sim
0.28, 0.52$ and 0.95 at $z=3.1$, $4.5$ and $6$, respectively. We are
using the \emph{model} luminosity function to compute these duty
cycles via equation~(\ref{eq|P0general}), but the agreement with the
observed $\Phi(L,z)$ is good enough (Figure~\ref{fig|r0z}) that we
can consider this a smooth proxy for the observational data.

\section{PREDICTIONS FOR $z>6$}
\label{sec|predictionszgt6}

The high duty cycle inferred at $z=6$ has profound implications for
the evolution of the luminosity function at still higher redshifts.
Between $z=3$ and $z=6$, the decreasing abundance of halos
with increasing redshift is partly compensated by the factor of
three increase in duty cycle. However, duty cycles cannot exceed
unity by definition, so at $z>6$ the fast drop of the massive and
rare host halos implies an equally rapid decline in the number
density of luminous quasars. At the same time, the implied mass of
quasar hosts moves even further out on the exponential tail of the
halo mass function. Our models thus predict a decline in high
redshift quasar numbers much steeper than expected from simple
extrapolations of the $z=3-6$ luminosity function.

\begin{figure}[ht!]
\epsscale{1.1} \plotone{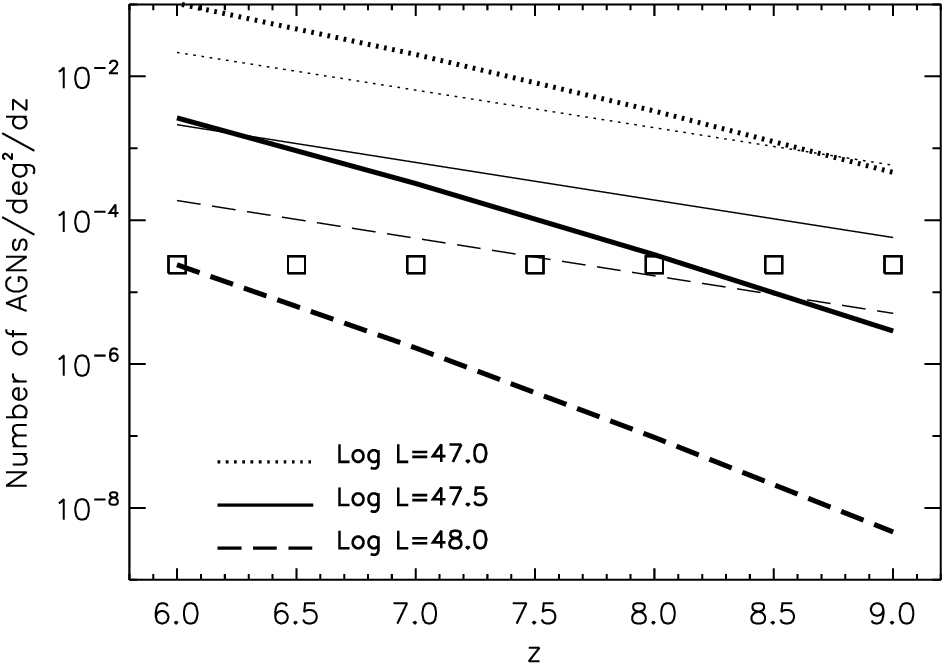}
\caption{Number counts of AGNs per square degree per unit redshift
as predicted by our reference model (\emph{thick} lines) as a
function of redshift, above the labeled luminosity thresholds.
\emph{Thin} lines refer to extrapolations of the Fan et al. (2004)
luminosity function at the same redshifts and luminosities. The
large \emph{open squares} indicate the number density per unit
redshift corresponding to one single observable quasar in the whole
sky.} \label{fig|Predictions}
\end{figure}

Figure~\ref{fig|Predictions} demonstrates this point, showing the
reference model predicted number counts of AGNs per square degree
per unit redshift as a function of redshift, above luminosity
thresholds of $\log L/{\rm erg\, s^{-1}}=47, 47.5,$ and $48$, as
labeled. Evolution is more rapid for higher luminosity AGN because
their host halos are further out on the tail of the mass function.
The thin lines refer to extrapolations of the Fan et al. (2004)
luminosity function at the same redshifts and luminosities. The
latter is a power law $\Phi(L)\propto L^{-3.1}$ that describes the
statistics of optical quasars in the range $\log L/{\rm erg\,
s^{-1}}\gtrsim 47$ and $5.5\lesssim z \lesssim 6.5$; we do not apply
any obscuration correction. Fan et al. (2004) find that a good
representation of the data requires redshift evolution
$\Phi(L,z)\propto 10^{-0.48\, z}$. It is evident from
Figure~\ref{fig|Predictions} that our reference model predicts a
decrease in AGN number density much faster than the one expected by
naively extrapolating the Fan et al. (2004) trend to $z\gtrsim 6$.
The open squares in Figure~\ref{fig|Predictions} indicate the number
density per unit redshift corresponding to one single observable
quasar in the whole sky. According to this model, the highest-$z$
quasar in the sky with true $L> 10^{47.5}\, {\rm erg\, s^{-1}}$
should be at $z \sim 7.5$ in our model; all quasars detected with
higher apparent luminosities by future surveys would have to be
magnified by lensing (see also Richards et al. 2004).

As recently discussed by Fontanot et al. (2007), the surface density
inferred from the luminosity function of Fan et al. (2004) and
Shankar \& Mathur (2007) predicts that only a few luminous sources
will be detected in the field of view of even the largest and
deepest future surveys such as JWST, and EUCLID. Our predictions
suggest that these detections will be even rarer than
simple empirical extrapolations predict.
While the predictions of Figure~\ref{fig|Predictions} are specific
to our adopted model parameters, this conclusion is likely to apply
more generally to models that reproduce the strong $z=4$ clustering
found by S07. This clustering implies high host halo masses and
hence high duty cycles at $z=4$, so the declining black hole mass
function cannot continue to be compensated by higher duty cycles
towards higher redshifts (though rapid evolution of the \mbh-\vvir
relation or rapidly increasing $\lambda$ values could compensate in
principle). Very similar results are found with the $\epsilon=0.1$
model of Figure~\ref{fig|r0z}.

Because the host halos of high redshift quasars are so highly
biased, the predicted clustering remains strong at $z>6$. For the
$\epsilon=0.15$ reference model, the predicted correlation length
$r_0$ as a function of $B$-band luminosity and redshift can be well
approximated by the relation $r_0(M_B,z)\simeq
27\times[(1+z)/7]^{0.3}(-26.5+M_B)^{0.5}$ Mpc, valid in the range
$-29\lesssim M_B \lesssim -26.5$ and $6\lesssim z\lesssim 9$.

\begin{figure}[ht!]
\epsscale{1.1} \plotone{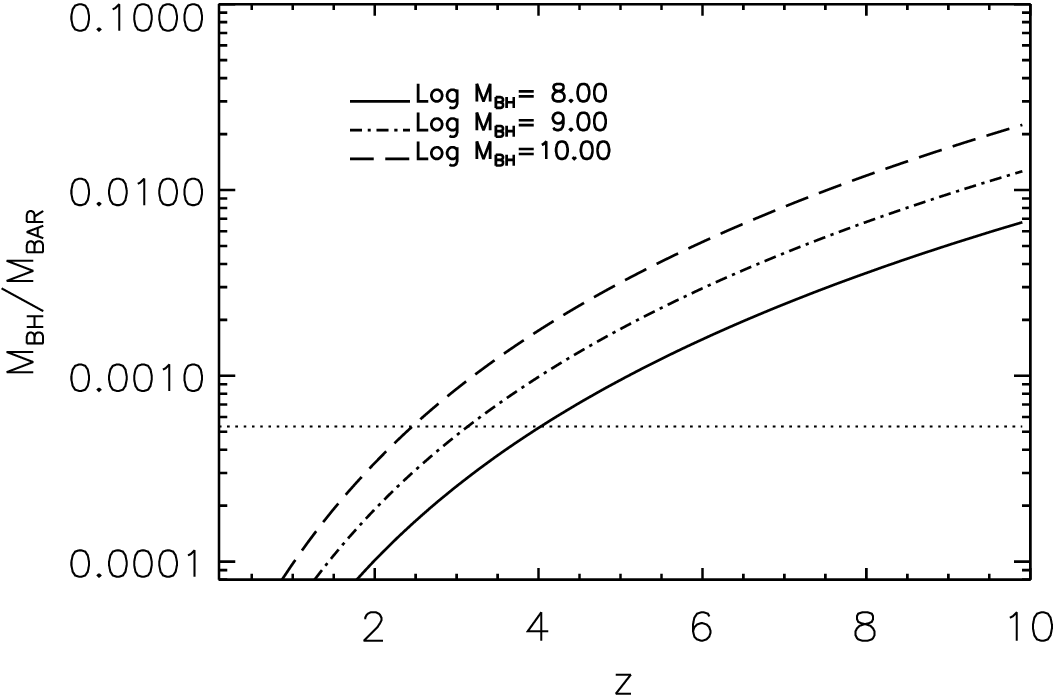} \caption{Ratio between black hole
and baryon mass within the halo, the latter computed as $M_{\rm
BAR}=0.17\times M$, for three values of the black hole mass, as
labeled. This ratio at $z\gtrsim 4$ gets higher than the local value
between black hole and bulge mass of $(1/3)\times 1.6\times 10^{-3}$
(\emph{dotted} line; see text), implying that at fixed stellar mass,
a larger fraction of the baryons in high mass halos is locked in the
central black hole at early times.} \label{fig|fbar}
\end{figure}

Local observations imply a ratio \mbh/$M_{\rm STAR}\approx 1.6\times
10^{-3}$ between the mass of the black hole and the stellar mass of
its host bulge (e.g., Marconi \& Hunt 2003, H\"{a}ring \& Rix 2004).
As discussed above, our reference model predicts increasing black
hole masses at fixed virial velocity ($\gamma>1$) at $z>3$ as
required to match the number density of very luminous quasars of
$z=6$. However, with the assumption that the baryon fraction within
a halo is universal, this implies that an increasing fraction of the
baryons must be locked up into the central black hole. We show this
in Figure~\ref{fig|fbar}, which plots the black hole-to-baryon
fraction as a function of redshift predicted by our reference model.
The baryonic mass fraction within any halo is set to be $M_{\rm
BAR}/M=f_b=0.17$ (e.g., Spergel et al. 2007; Crain et al. 2007).
Given that the stellar mass $M_{\rm STAR}$ in the local universe is
unlikely to exceed $f_b$, and is more typically $\lesssim f_b/3$ in
massive galaxies (e. g., Shankar et al. 2006), this implies that the
black hole mass is $\lesssim 1.6\times 10^{-3}/3$ the mass of the
total baryons in the host halo. This local ratio between black hole
and baryonic mass is shown as a horizontal dotted line in
Figure~\ref{fig|fbar}. The fraction of baryons locked in the central
black hole increases at higher redshifts following the increase of
virial velocity at fixed halo mass (Eq.~[\ref{eq|Vvir}]) and the
increase of black hole mass at fixed virial velocity proportional to
$(1+z)^{\gamma}$ (Eq~[\ref{eq|MbhVvirRelation}]). For our reference
model, the \mbh/$M_{\rm BAR}$ ratio grows rapidly at high redshifts
and exceeds the local value by nearly an order of magnitude at
$z\gtrsim 6$ for \mbh$\ge 10^9\, {\rm M_{\odot}}$. Note that even a
model with $\gamma=0$ still produces an increase of the $M_{\rm
BAR}/M$ ratio with redshift, driven by the redshift dependence in
equation~(\ref{eq|Vvir}), although it is just a factor of a few in
this case.

Therefore, we conclude that the relation between black hole and
spheroidal stellar mass determined locally cannot continue to hold
at very high redshifts if the large clustering strength reported at
$z=4$ is to be matched, and that a much larger fraction of baryons
in galaxies must accrete to the nuclear black holes at $z\gtrsim 4$.

\section{COMPARISON TO CONSTANT DUTY CYCLE MODELS}
\label{sec|soltan}

The results in \S~\ref{sec|results} show that matching the high
clustering signal measured by S07 requires a high duty cycle $P_0$,
which corresponds to quasars preferentially residing in high mass,
less abundant halos. This result has also been discussed by S07 and
by WMC, following the method outlined by Martini \& Weinberg (2001)
and Haiman \& Hui (2001). The model described in \S~\ref{sec|method}
assumes an \emph{a priori} relation between luminosity $L$ and halo
mass $M$. Since this model also predicts the AGN luminosity function
from the equation governing the growth of black holes, it implicitly
predicts the duty cycle required to assign an AGN luminosity to a
halo mass and match their abundances. Martini \& Weinberg (2001) and
WMC instead define the relation between $L$ and $M$ \emph{a
posteriori}, i.e., from the cumulative matching between the
\emph{observed} AGN luminosity function and the halo mass function,
once an input duty cycle has been specified. Since both methods
assume a (nearly) monotonic relation between luminosity and halo
mass, they should yield a similar connection of duty cycle and
clustering in cases where the \emph{a priori} model matches the
observed luminosity function.

\begin{figure*}[ht!]
\epsscale{1.1} \plotone{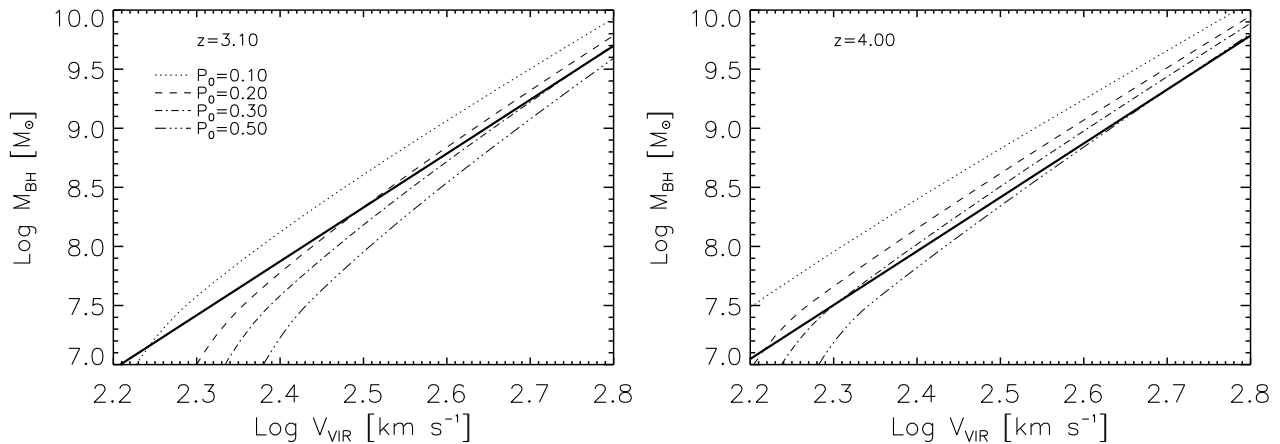}
\caption{Correlation between black hole mass and halo virial
velocity implied by the cumulative matching between the halo and
black hole mass functions, the latter derived from the reference
model luminosity function and a constant input duty cycle $P_0$. The
different lines refer to different values of the duty cycle, as
labeled, while the \emph{solid} line is the \mbh-\vvir relation
corresponding to our reference model. The left and right panels show
the resulting relations at redshifts $z=3.1$ and $z=4$,
respectively.} \label{fig|MbhVvir}
\end{figure*}

To compare the two approaches in detail, we compute the relation
between black hole mass and virial velocity for fixed duty cycle via
the equation
\begin{equation}
\Phi_{\rm BH}(>M_{\rm BH},z)=\frac{\Phi(>L,z)}{P_0}=\Phi_h(>V_{\rm
vir},z)\, ,
    \label{eq|NumberMatching}
\end{equation}
where $\Phi(>L,z)$ is the model predicted
AGN luminosity function, and $\Phi_h(V_{\rm vir},z)$ is derived from the halo mass function
and the \vvir-\mh relation of equation~(\ref{eq|Vvir}). We assume
$\lambda=0.5$ to convert from \mbh$\,$ to $L$.
Figure~\ref{fig|MbhVvir} plots the relations implied by
equation~(\ref{eq|NumberMatching}) at redshifts $z=3.1$ and $z=4$.
Curves from top to bottom show the \mbh-\vvir relation assuming a
constant duty cycle $P_0=0.1, 0.2, 0.3$ and $0.5$. Higher $P_0$
corresponds to rarer halos, hence higher \vvir and stronger
clustering. The solid curves in the two panels represent the output
\mbh-\vvir relation for our reference model, which predicts a duty
cycle of $P_0(z=3.1)\sim 0.30$ and $P_0(z=4.0)\sim 0.50$ at the high
mass end. As expected, the \mbh-\vvir relation of our reference
model is similar at high masses to that model with similar duty
cycle. At lower masses, our model does not perfectly match the
observed luminosity function and does not predict a constant duty
cycle.

\begin{figure*}[ht!]
\epsscale{1.0} \plotone{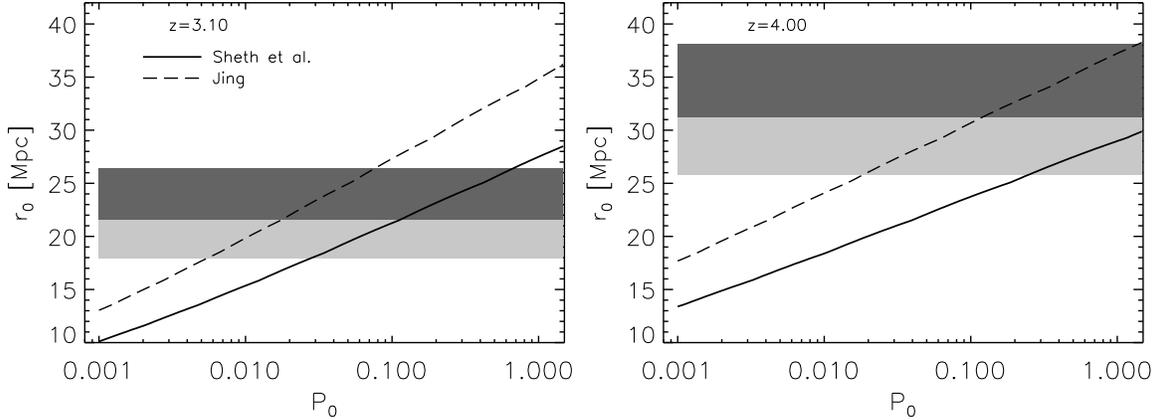} \caption{Predicted clustering
correlation length $r_0$ computed above the minimum survey
sensitivity as a function of duty cycle and adopting the luminosity
function derived from our reference model. The \emph{solid} and
\emph{long-dashed} lines refer to the $r_0$ implied by using Jing's
(1998) and Sheth et al.'s (2001) bias formula, respectively. The
left and right panels show our results at redshifts $z=3.1$ and
$z=4$, respectively. Dark and light shaded bands show the $1\sigma$
range of the S07 measurements at these redshifts, from ``good
fields'' and ``all fields'', respectively. Maximal values of the
duty cycle predict a clustering strength only marginally consistent
with the data at $z=4$.} \label{fig|r0zDuty}
\end{figure*}

We compute the average bias for the constant duty cycle model via
\begin{equation}
\bar{b}(z)=\frac{\int_{M_{\rm
min}(z)}^{\infty}b(M,z)\Phi_h(M,z)d\log M}{\int_{M_{\rm
min}(z)}^{\infty}\Phi_h(M,z)d\log M}\, ,
    \label{eq|biasCum}
\end{equation}
where $M_{\rm min}(z)$ is the halo mass corresponding to $L_{\rm
min}(z)$ via equations~(\ref{eq|Vvir}) and
(\ref{eq|NumberMatching}). Because the relation between $L$ and
\vvir is determined by matching space densities, the predicted bias
is independent of $\lambda$. Figure~\ref{fig|r0zDuty} plots the
corresponding correlation length $r_0$, computed through
equation~(\ref{eq|r0}), as a function of $P_0$ at $z=3.1$ and $4.0$.
Solid and dashed lines show the results of using the Sheth et al.
(2001) and Jing (1998) formulas, respectively, for the bias $b(M,z)$
in equation~(\ref{eq|biasCum}). Shaded regions show the $1\sigma$
range of the S07 measurements. As expected the clustering strength
strongly increases with increasing duty cycle. In agreement with
WMC, we find that a duty cycle $P_0\gtrsim 0.2$ is required to
reproduce S07's $z=4$ measurement with the Jing (1998) bias formula.
However, our N-body results favor the Sheth et al. (2001) bias
formula, and in this case we cannot match the S07 ``good''
measurement within $1\sigma$ at $z=4$ even for a maximal duty cycle
$P_0=1$.

\begin{figure*}[ht!]
\epsscale{1.0} \plotone{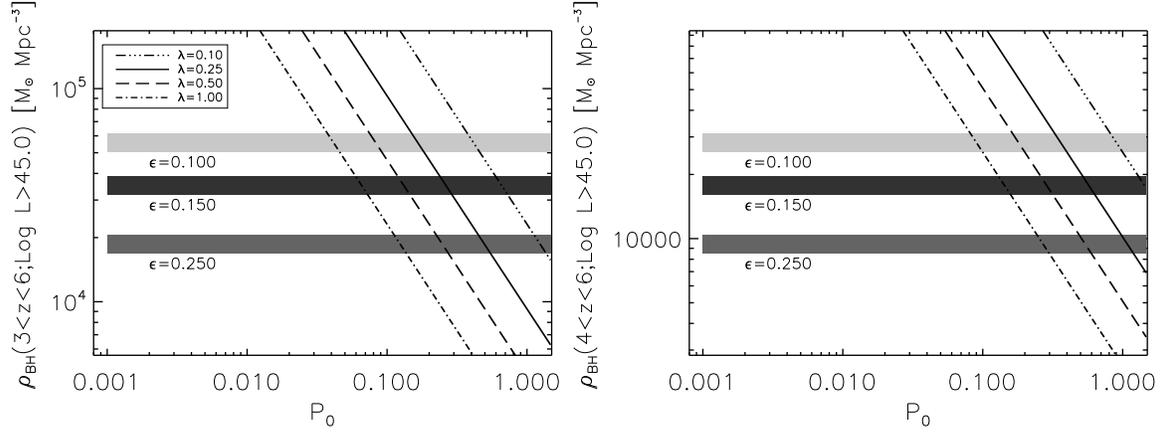} \caption{\emph{Left} panel: the
\emph{horizontal stripes} show the integrated black hole mass
density above $L=10^{45}\, {\rm erg\, s^{-1}}$ and $3.1<z<6$ derived
from our reference model luminosity function for three different
values of the radiative efficiency, as labeled; \emph{triple dot-
dashed}, \emph{solid}, \emph{long-dashed}, and \emph{dot-dashed}
lines indicate the black hole mass density implied by the luminosity
function at $z=3.1$ integrated over mass assuming a mean Eddington
ratio $\lambda=0.1,0.25,0.5,1$, respectively, as a function of duty
cycle $P_0$. \emph{Right} panel: same as left panel but integrating
the luminosity function over the redshift range $4<z<6$. For
acceptable combinations of $\lambda$, $\epsilon$, and $P_0$, the
line corresponding to $\lambda$ should intersect the band
corresponding to $\epsilon$ at duty cycle $P_0$. High duty cycles
require high radiative efficiencies or low Eddington ratios to
reconcile the cumulative AGN emissivity with the black hole mass
density in rare halos (see text).} \label{fig|Soltan}
\end{figure*}

As shown in \S~\ref{sec|results}, models with high duty cycles
require high $f/\lambda$ ratios to reproduce the observed luminosity
function. This connection can be simply understood from
equation~(\ref{eq|continEqInv}): increasing the duty cycle decreases
the number density of halos that host AGNs, which in turn need to
increase their \emph{e}-folding time $t_{\rm ef}$ to maintain the
same observed luminosity density. Figure~\ref{fig|Soltan} shows this
effect in detail. We first integrate our reference model luminosity
function from $z=6$ down to a given redshift $z$ as
\begin{equation}
\rho_{\rm BH}(>\log L,z)=\frac{1-\epsilon}{\epsilon c^2}\int_{z}^6
dz'\int_{\log
L}^{\infty}\Phi(L',z')L'\left|\frac{dt}{dz'}\right|d\log L'\, .
    \label{eq|soltan}
\end{equation}
We consider only luminous AGNs that shine with luminosity $\log
L/{\rm erg\, s^{-1}}\ge 45$, corresponding to black hole masses
above \mbh$\sim 10^7\,M_{\odot}$, which ensures that we are properly
tracking the accretion histories of the most massive black holes. It
is evident from equation~(\ref{eq|soltan}) that the accreted mass
density does not depend on the black hole Eddington ratio
distribution but only on the radiative efficiency. Our results are
shown as stripes in Figure~\ref{fig|Soltan} for three different
values of the radiative efficiency $\epsilon=0.10, 0.15$ and $0.25$,
from top to bottom. The left and right panels of
Figure~\ref{fig|Soltan} show the integrated mass density at $z=3.1$
and $4$, respectively. Note that these are the black hole mass
densities implied by the So{\l}tan (1982) argument given an input
luminosity function that is a good match to observations.

Alternatively, by assuming an average duty cycle $P_0(z)$ at a given
redshift $z$ we can convert the AGN luminosity function into a black
hole mass density via
\begin{equation}
\rho'_{\rm BH}(>\log L,z)=\int_{\log L}^{\infty} \frac{L'}{\lambda
P_0 l}\Phi(L',z)d\log L'\, .
    \label{eq|rhoBHatgivenz}
\end{equation}
The latter estimate\footnote{Note that we do not consider any
scatter between black hole mass and AGN luminosity in
equation~(\ref{eq|rhoBHatgivenz}), as the luminosity function has
been derived from the continuity equation in
equation~(\ref{eq|continEqInv}), which requires a strictly monotonic
relation between black hole and halo mass.} depends inversely on the
Eddington ratio $\lambda$ because of the mapping of $L$ to \mbh, but
it does not depend on the radiative efficiency. We plot $\rho'_{\rm
BH}$ as a function of the duty cycle $P_0$ for $\lambda=0.25$, 0.5,
and 1.0, with solid, dashed, and dotted lines, respectively. Results
for the $z=3.1$ and $z=4$ accreted mass density are shown in the
left and right panels of Figure~\ref{fig|Soltan}, respectively. It
is noteworthy that the high radiative efficiency of $\epsilon=0.15$,
as used in our reference model, is consistent with $P_0(z=3.1)\sim
0.28$ and $P_0(z=4)\sim 0.50$, in perfect agreement with our
findings presented in Figure~\ref{fig|dutycycle}.

Overall, we find evidence for a general rule of thumb: if black
holes accrete at a significant fraction of the Eddington luminosity
($\lambda \gtrsim 0.25$) and possess high duty cycles as derived
from their strong clustering (Figure~\ref{fig|r0zDuty}), then they
must also radiate at \emph{high} radiative efficiencies
($\epsilon\gtrsim 0.15$) to match the AGN luminosity function and
its evolution with redshift. This conclusion from constant duty
cycle models is entirely consistent with our conclusions from
\mbh-\vvir models discussed
in \S\S~\ref{sec|method}-\ref{sec|predictionszgt6}.

\section{DISCUSSION AND CONCLUSIONS}
\label{sec|conclu}

We have investigated constraints on the host halos, radiative
efficiencies and active duty cycles of high redshift black holes
that are implied by recent measurements of the AGN luminosity
function at $3\le z \le 6$ and of optical quasar clustering at
$z\approx 3$ and $z\approx 4$. In this work we have derived the
predicted AGN luminosity function implied by a model black hole mass
function. The latter is built from the dark matter halo mass
function at each redshift by applying a model relation between black
hole mass and halo virial velocity, motivated by local observations.
Our models are parameterized by the high-redshift normalization
$\alpha$ and redshift evolution index $\gamma$ of the mean
\mbh-\vvir relation (equation~[\ref{eq|MbhVvirRelation}]), by the
log-normal scatter $\Sigma$ about this relation (in dex), and by the
Eddington ratio $\lambda$ and radiative efficiency $\epsilon$ of
black hole accretion.

A reference model with $(\alpha, \gamma, \Sigma, \lambda,
\epsilon)=(1.1, 1.0,0.1,0.25,0.15)$ provides a good fit to the $z=3$
correlation length $r_0$ and a reasonable fit to the bright end of
the luminosity function ($L\gtrsim 10^{46.5}\, {\rm erg\, s^{-1}}$)
at $z=3-6$. It overpredicts the faint end of the luminosity
function, probably indicating that our assumption of a constant
$\lambda$ or power-law \mbh-\vvir relation breaks down in this
regime. More significantly, the model prediction is below S07's
estimate of $r_0$ for luminous quasars at $z=4$, by about $2\sigma$.
While lowering $\alpha$ or $\lambda$ raises the predicted $r_0$, it
lowers the predicted luminosity function below the observations,
unless we allow efficiencies greater than $\epsilon=0.15$.
Increasing the scatter $\Sigma$ reduces the predicted clustering,
making the overall fit to the data worse.  If we use S07's ``all
field'' estimate of $r_0$ instead of their ``good field'' estimate,
then the discrepancy at $z=4$ is under $1\sigma$.  The reference
model predicts substantial luminosity and redshift dependence of the
quasar correlation length at $z>3$ (Figure~\ref{fig|bLz}), with $r_0
\approx 27\times[(1+z)/7]^{0.3}(-26.5+M_B)^{0.5}$ Mpc for $6 \leq z
\leq 9$.

Models that successfully match the high redshift bias at $z\gtrsim
3$ require luminous AGNs to reside in massive and therefore rare
halos, implying high duty cycles, $P_0 \sim 0.2,$ 0.5, 0.9 at
$z=3.1$, 4.5, 6.0 in our reference model. Note that, although this
model is consistent with the $z=4$ clustering only at the $2\sigma$
level, it already produces a duty cycle close to unity at high
redshifts. Raising the predicted correlation requires putting
quasars in more massive, less numerous halos, and thus tends to push
the required duty cycle above unity.

To simultaneously reproduce the observed luminosity function and
bias, models must have $f/\lambda \gtrsim 0.7$, where $f =
\epsilon/(1-\epsilon)$, so that the mass density of black holes in
these rare halos corresponds to the cumulative emissivity of the
luminous AGN. These findings are robust against uncertainties in the
obscured fraction of AGNs or in the precise value of the mean
bolometric correction (see discussion in \S~\ref{subsubsec|chi2}).
The underlying physics that leads to these findings is easy to
understand.  The strong observed clustering at $z=4$ implies a high
duty cycle and thus a low space density of massive black holes.
Reproducing the observed AGN emissivity with the low total mass
density in black holes requires a high radiative efficiency.
Lowering the assumed Eddington ratio implies a higher mass density (because
each black hole is more massive) and a proportionally lower $f$.
As shown in the Appendices, mergers are expected to have little
impact on the BH mass function at these redshifts, but
to the extent they do have an impact they raise the
limit on $f/\lambda$ by adding mass without associated luminosity.

For any choice of the mean Eddington ratio, our successful models
require positive evolution of the \mbh-\vvir relation ($\gamma > 0$)
at $z>3$ to reproduce the evolving bright end of the luminosity
function. Evolution of the Eddington ratio itself (higher $\lambda$
at higher $z$) could in principle yield similar evolution.

It is beyond the purposes of the current
work to extrapolate the simple model
outlined here to lower redshifts. First of all, the basic
treatment presented by SWM has shown that the large scale
clustering of quasars can be simply matched by
accretion models which evolve the black hole mass function
assuming reasonable values
of the radiative efficiency and Eddington ratios,
which satisfy Soltan (1982)'s constraint. Moreover,
at lower redshifts, several additional physical inputs
need to be added to the model (e.g., the fraction
of active satellites, mass-dependent Eddington ratios, AGN feedback) to
reproduce the full quasar clustering at all luminosities, scales, and redshifts (e.g., Wyithe \& Loeb 2003; Scannapieco \& Ho 2004; Hopkins et al. 2008; Bonoli et al. 2009a; Thacker et al. 2009).

Previous work that attempted to simultaneously match the quasar
luminosity function and bias has yielded somewhat different results
from ours. Wyithe \& Loeb (2003, 2005; see also Rhook \& Haehnelt
2006) developed a model aimed at reproducing both the bias and the
AGN luminosity function at several redshifts. They expressed the
relation between the luminosity and halo mass via some AGN
feedback-motivated models for the black hole-halo relation, and they
assumed that black holes grow at the Eddington limit and radiative
efficiency of $\epsilon=0.1$. Their values of $f/\lambda$ would then
be lower by a factor of several with respect to ours. These
differences are due to a different AGN bolometric luminosity
function used (ours being a factor of a few higher) and the absence
of the SDSS bias measurements at $z>3$. In brief, we do not think
that these models would reproduce the observational data considered
in this paper.

Our lower limit on $f/\lambda$ translates to a lower limit on
radiative efficiency $\epsilon \ga 0.7\lambda/(1+0.7\lambda)$.  With
observationally estimated values $\lambda \approx 0.3$ for the
Eddington ratios of luminous high-redshift quasars (Kollmeier et
al.\ 2006; Shen et al.\ 2008), this limit implies $\epsilon \ga
0.17$. Using a different approach that links the observed AGN
luminosity function to the local black hole mass function via the
continuity equation, a differential generalization of So{\l}tan's
(1982) cumulative emissivity argument, SWM estimate an average
radiative efficiency $\epsilon \approx 0.05-0.10$.  Other authors
pursuing similar approaches and adopting similar bolometric
corrections have reached similar conclusions (e.g., Salucci et al.
1999; Marconi et al. 2004; Shankar et al. 2004; Hopkins et al.
2007a; Yu \& Lu 2008; see SWM for further discussion). As discussed
in detail by SWM, uncertainties in the local black hole mass
function, bolometric corrections, and obscured fractions still leave
significant range in the inferred value of $\epsilon$, but these
uncertainties would have to be pushed to their extremes to
accommodate $\epsilon \ga 0.17$.

One possible resolution of this tension is that the typical
radiative efficiency is higher at $z>3$ than it is at the lower
redshifts that dominate the overall growth of the black hole
population.  However, the similarity of quasar spectral energy
distributions at low and high redshifts (e.g., Richards et al.
2006b) argues against a systematic change in accretion physics.  We
should therefore examine the loopholes in the argument for high
efficiency presented here, noting that it is above all the $z=4$
clustering measurement from S07 that drives our models to massive,
rare halos and thus to high efficiencies to reproduce the luminosity
function. Adopting the Jing (1998) bias formula instead of the Sheth
et al. (2001) formula would allow us to match the clustering with
less massive halos, but our numerical simulation results show that
the Sheth et al. (2001) formula is more accurate in the relevant
range of halo mass and redshift.

Our conclusion is insensitive to the specific assumption of a
one-to-one power-law relation between black hole mass and halo
virial velocity: monotonically matching luminosity functions and
halo mass functions leads to a similar conclusion
(\S\ref{sec|soltan}; WMC), and adding scatter, while plausible on
physical grounds, only reduces clustering and thus exacerbates the
underlying tension.  Because the implied characteristic halo mass is
already well above $M_*$, halos hosting two or more quasars should
be far too rare to significantly alter the large scale bias. Small-scale clustering studies of large $z>3$ quasar samples (Hennawi et al. 2009; Shen et al. 2009), will set more definite constraints
on the actual fraction of active subhalos.

Our modeling does assume that halos hosting quasars have the same bias
as average halos of the same mass, while Wechsler et al.\ (2006)
find that the youngest 25\% of high redshift, high $M/M_*$ halos
have a correlation length $\sim 13\%$ higher than the average
correlation length of halos at the same mass and redshift.
Clustering could be slightly boosted if active quasars
preferentially occupy these younger halos, but the high duty cycle
required in our models effectively closes this loophole, implying
that the quasar host population includes the majority of massive
halos rather than a small subset. Wyithe \& Loeb (2009) suggest that
the strong $z=4$ clustering could be explained by tying quasar
activity to recently merged halos, which might have stronger bias
(see also Furlanetto \& Kamionkowski 2006; Wetzel, Cohn, \& White 2009).
However, we suspect that halos with substantial
excess bias might be too rare
to satisfy duty cycle constraints,
and a recent study by Bonoli et al. (2009b) uses outputs from the
Millennium Simulation (Springel et al. 2005) to show that no excess
bias is present in recently merged massive halos.

Perhaps the most plausible loophole in our conclusion is simply that
the S07 $z=4$ correlation length is overestimated (or its
statistical error underestimated), since it is the first measurement
of its kind and there is a significant difference between the S07
values for ``all fields'' and ``good fields'' (even though only 15\%
of fields are excluded from the latter sample). Even the central
value of the (lower) ``all fields'' measurement requires high
$\epsilon$ or low $\lambda$ when combined with luminosity function
constraints, but the lower $1\sigma$ value can be reconciled with
$\epsilon \approx 0.1$ and $\lambda \approx 0.25$. The DR7 SDSS
quasar sample should afford substantially better statistics than the
DR5 sample analyzed by S07, allowing stronger conclusions about the
host halo population.

In these models, the rapid decline in the number of luminous quasars
between $z=3$ and $z=6$ is driven by the rapidly declining abundance
of halos massive enough to host them.  However, the drop in halo
abundance is partly compensated by a rise in the duty cycle over
this interval, from $\sim 0.2$ to $\sim 0.9$ in our reference model.
Since duty cycles cannot exceed one by definition, this compensation
cannot continue much beyond $z=6$, and the decline in host halo
abundance accelerates because these halos are far out on the
exponential tail of the mass function.  The predicted evolution of
the quasar population at $z>6$ is therefore much more rapid than
simple extrapolations of the observed $z=3-6$ behavior.  This break
to more rapid evolution at $z>6$ should be a generic prediction of
models that reproduce the strong observed clustering at $z=4$,
though in principle it could be softened by a rapid increase of
Eddington ratios at $z>6$ or by a sudden change in evolution of the
\mbh-\vvir relation.  Surveys from the next generation of wide-field
infrared instruments will have to probe to low luminosities to
reveal the population of growing black holes at $z>7$.


\begin{acknowledgements}
We thank Yue Shen and Paul Martini for useful discussions and
comments. This work was supported by NASA Grant NNG05GH77G. MC
acknowledges support from the Spanish Ministerio de Ciencia y
Tegnologia (MEC) through a Juan de la Cierva program,  and grant
AYA2006-06341. PF acknowledges support from the Spanish MEC through
a Ramon y Cajal fellowship, and grants AYA2006-06341, 2005SGR-00728.
We acknowledge the use of simulations from the MICE Consortium
(www.ice.cat/mice) developed at the MareNostrum supercomputer
(www.bsc.es) and stored at the PIC (www.pic.es).
\end{acknowledgements}

\begin{appendix}
\section{Halo Bias at high redshift}

In this Appendix we provide additional details of our bias analysis
from the N-body simulation introduced in
\S~\ref{subsec|comparesimul}, in particular the difference between
using quasi-linear scales (e.g. $8-38\,\,h^{-1}\,{\rm Mpc}$ as
reported) or larger ones. We also comment on the distinction between
deriving the bias from the halo autocorrelation $\xi_{hh}$ or the
halo-matter correlation function $\xi_{hm}$. We finish by comparing
our results with those of S07.

The simulation used was provided by the MICE collaboration (Fosalba
et al. 2008) and contains $1024^3$ particles in a cubic volume of
side $L_{\rm box}=768\,h^{-1}\,{\rm Mpc}$, with cosmological
parameters $\Omega_m=0.25$, $\Omega_L=0.75$, $\sigma_8=0.8$,
$n=0.95$, $h\equiv H_0/100\, {\rm km\, s^{-1}\, Mpc^{-1}}=0.7$ and
$\Omega_b=0.044$. Halos were identified in the comoving outputs at
$z=3,4,5,5.5$ and $6$ using the friends-of-friends algorithm with
linking length $b=0.164$. Finally, at each redshift the
corresponding halo catalogue was divided in three non-overlapping
sub-samples of halo masses in the ranges $[5.6-17.7]\times 10^{11}
M_{\odot}$, $[1.7-5.6]\times 10^{12} M_{\odot}$ and
$[5.6-17.7]\times 10^{12} M_{\odot}$ respectively (i.e. bins of
equal width in $\log M$).

\begin{figure}
\epsscale{1.0} \plotone{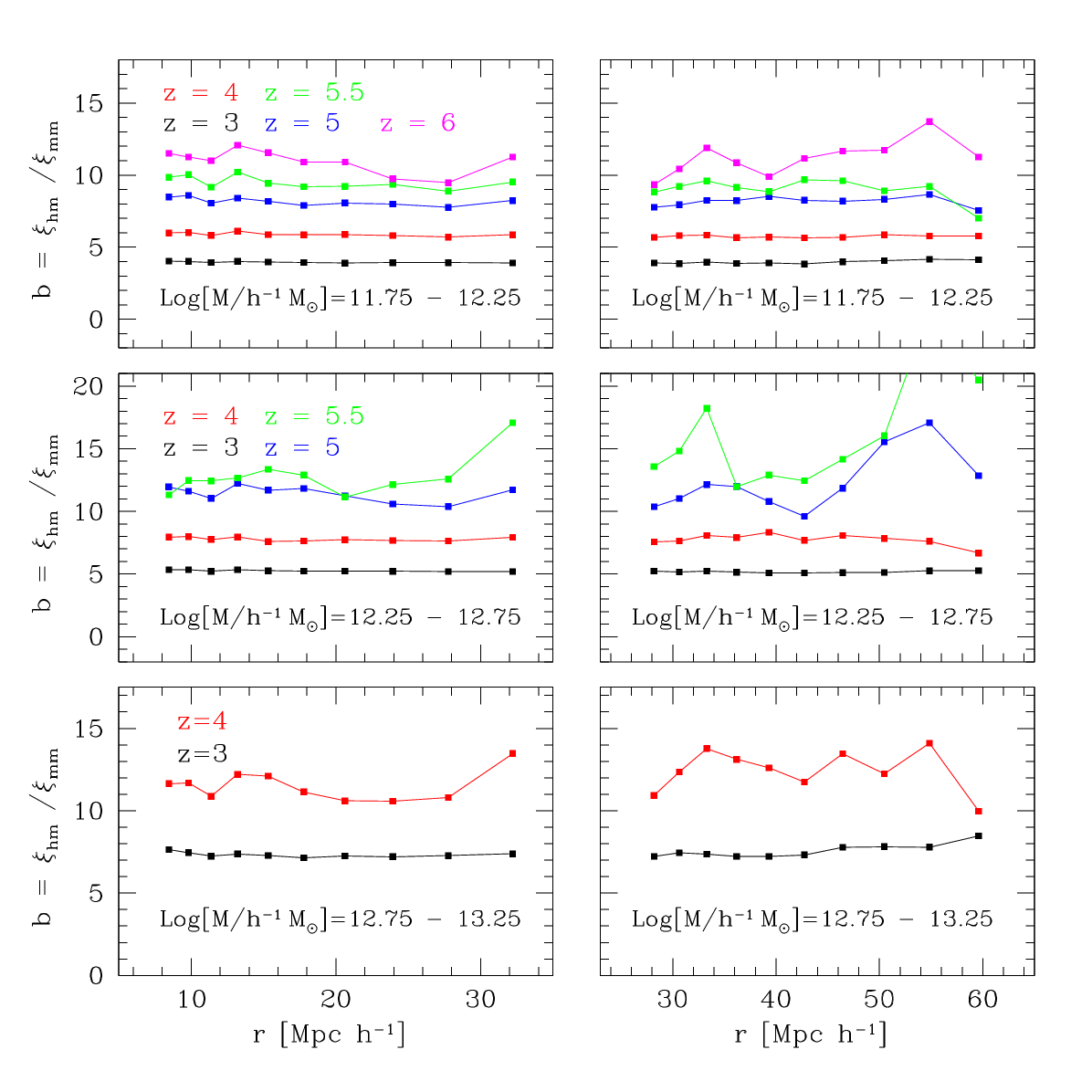} \caption{Halo bias as a function
of scale. \emph{Right} panels: the ratio $\xi_{hm}/\xi_{mm}$ at
large scales for different halo masses and redshifts, as labeled.
\emph{Left} panels: the same ratio at smaller separations, that
encompass the scales where AGN clustering has being measured by S07.
At smaller separations there is not significant scale dependence
while the scatter of the measurement is lower than for large
separations. The measured bias is higher by $2-3\,\%$ in the left
hand panels, but this difference is within the scatter.}
\label{fig|Small.vs.Large}
\end{figure}

We have obtained the halo bias from the ratio of correlation
functions $\xi_{hm}(r)/\xi_{mm}(r)$ averaged over $10$ bins of equal
length in $\log r$ in the radial range $8\,h^{-1}\,{\rm Mpc}\le r
\le 38\,h^{-1}\,{\rm Mpc}$. We also implemented the same
measurements in the radial range $28\,h^{-1}\,{\rm Mpc}\le r \le
60\,h^{-1}\,{\rm Mpc}$ in order to test for any dependence of the
bias on scale. In Fig.~\ref{fig|Small.vs.Large} we show the ratio
$\xi_{hm}/\xi_{mm}$ at both ranges for all redshifts and mass bins
studied.  On the one hand, this figure indicates that within the
intrinsic scatter there is no significant scale dependence of the
bias at smaller separations. On the other hand, the values of the
measured bias rise by only $2-3\,\%$ when using
$8-38\,\,h^{-1}\,{\rm Mpc}$ instead of $28-60\,\,h^{-1}\,{\rm Mpc}$,
but this difference is within the variance of the simulation.

In order to overcome the shot noise due to the low abundance of rare
halos we decided to measure the bias from cross correlating the halo
distribution to that of the dark matter. This allowed us to extend
the measurements to cases where the halo-halo correlation is too
noisy to define a meaningful bias. In Table 1 we report the bias
results for $\sqrt{\xi_{hh}/\xi_{mm}}$ (left table) and for
$\xi_{hm}/\xi_{mm}$ (right table). In both cases we measured at
scales in the range $8\,h^{-1}\,{\rm Mpc}\le r \le 38\,h^{-1}\,{\rm
Mpc}$. The reported error corresponds to the variance among
different bins (which might be taken as a rough representation of
the true error of the measurement with the caveat that different
bins are correlated). We find consistent results for the values of
the bias derived from the two methods in those bins of mass and
redshift where a reliable estimate can be obtained.

\begin{table}[ht!]
\begin{center}
\begin{tabular}{|c|c|c|c|}
\hline
$\log({\rm M}/{\rm M}_{\odot})=$    & $ 11.75 - 12.25$    & $12.25 - 12.75$    & $12.75-13.25  $  \\
                                    &                     &                    &                  \\
$z=3$                                & $ b=3.95 \pm 0.07 $ & $
b=5.3\pm 0.17 $  & $ b=7.8 \pm 0.69$\\ \hline z=4
& $ b=5.97 \pm 0.22 $ & $ b=8.1 \pm 0.57 $ &                  \\
\hline $z=5$                                 & $ b=8.44 \pm 0.61 $ &
$b=11.6\pm 2.78 $  &                  \\ \hline z=5.5
& $ b=10.1 \pm 1.1  $ &                    &                  \\
\hline $z=6$                                 & $ b=12.3 \pm 1.6  $ &
&                  \\ \hline
\end{tabular}
\begin{tabular}{|c|c|c|c|}
\hline
$\log({\rm M}/{\rm M}_{\odot})=$    & $ 11.75 - 12.25$    & $12.25 - 12.75$     & $12.75-13.25  $   \\
                                    &                     &                     &                   \\
$z=3$                                 & $ b=3.97 \pm 0.06 $ & $
b=5.27\pm 0.09 $  & $ b=7.38 \pm 0.2$ \\ \hline z=4
& $ b=5.9  \pm 0.13 $ & $ b=7.88 \pm 0.22 $ & $ b=11.52\pm0.54$ \\
\hline $z=5$                                 & $ b=8.25 \pm 0.27 $ &
$b=11.44\pm 0.5  $  &                   \\ \hline z=5.5
& $ b=9.53 \pm 0.39 $ & $ b=12.78\pm 1.4$   &                   \\
\hline $z=6$                                 & $ b=10.96\pm 0.69 $ &
&                   \\ \hline
\end{tabular}
\end{center}
\caption{Halo bias obtained from $b=\sqrt{\xi_{hh}/\xi_{mm}}$
(\emph{left} table) or $b=\xi_{hm}/\xi_{mm}$ (\emph{right} table).
The values are consistent with each other within the bin-to-bin
scatter, which is reported as the corresponding error.}
\label{tab|bias}
\end{table}

Finally, we  compare our results to those of S07, who obtained the
halo bias from $b=\sqrt{\xi^{hh}_{20}/\xi^{mm}_{20}}$ with,
\begin{equation}
\xi_{20}=\frac{3}{r^3_{\rm max}}\int_{r_{\rm min}}^{r_{\rm max}}
\xi(r) r^2 dr\, ,
\end{equation}
where $r_{\rm min}=5\,h^{-1}\,{\rm Mpc}$ and $r_{\rm
max}=20\,h^{-1}{\rm Mpc}$. Using all halos more massive than $M_{\rm
min}=2 \times 10^{12}\,h^{-1}\,M_{\odot}$, S07 finds $b_{\rm
eff}(z=3)=6.2$ and $b_{\rm eff}(z=4)=10.2$ (respectively $17\%$ and
$5\%$ lower than Jing's [1998] prediction). Using a similar mass-cut
to S07 and measuring the bias in the same way (i.e. from $\xi_{20}$)
we obtain $b_{\rm eff}(z=3)= 6.07$ and $b_{\rm eff}(z=4)= 9.35$
(where we have included a $6\%$ correction due to the difference in
the assumed cosmology). These values are in good agreement (within
$8\%$). However, and contrary to S07, we have chosen Sheth, Mo \&
Tormen (2001) expression as our primary bias prediction for reasons
already outlined in \S~\ref{subsec|comparesimul}.

\section{The contribution of BH mergers to mass accretion}

In this Appendix we compute the expected contribution
from mergers to the overall mass growth of the central black hole
of a halo with mass $M_0\sim 1-2\times 10^{13}$\msun\ at $z\sim 3-4$,
typical of the $z>3$ quasar hosts studied in this paper.

\begin{figure}
\epsscale{1.0} \plottwo{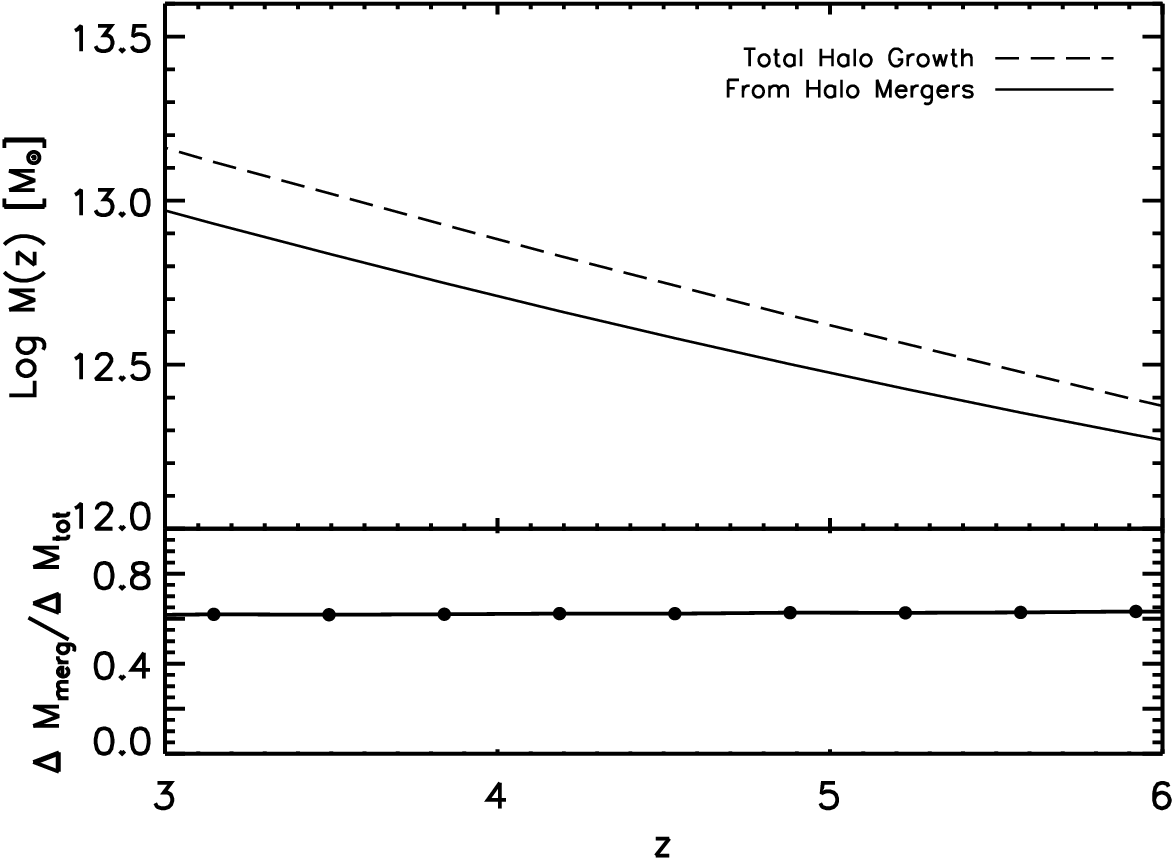}{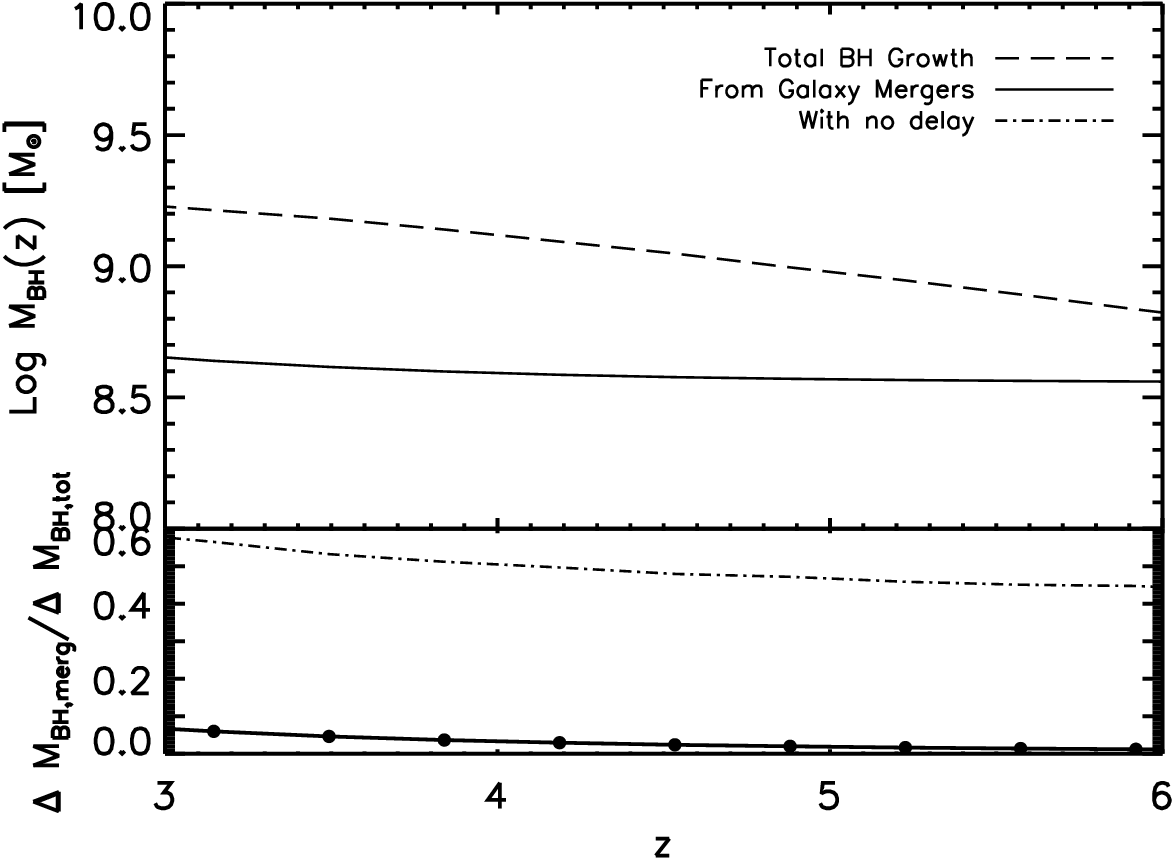} \caption{\emph{Left}: Growth history
of halos with final mass $\sim 1-2\times 10^{13}$\msun, as derived from
the halo mass function (\emph{long-dashed} line), compared to the growth
history as predicted by integrating the merger rates of halos with
progenitors mass ratios in the range $0.1<\xi<1$ (\emph{solid} line). \emph{Right}: the \emph{long-dashed} line shows the total growth in the central black hole as derived from the long-dashed line in the left panel and using equation~(\ref{eq|MbhMhalo}); the \emph{solid} line is instead the black hole growth derived by integrating the central galaxy merger rates. The \emph{dot-dashed} line
in the lower panel corresponds to black hole growth
when no delay is considered. Mergers contribute by only $\lesssim 7\%$ to the overall growth of the central black hole, at the most, if the delay due to dynamical friction is taken into account. See text for details.}
\label{fig|MergersAppendix}
\end{figure}

We trace the average mass growth history $M(z)$
of a such halos at any redshift $z$ by
imposing the number density conservation within a comoving volume
\begin{equation}
n[M(z),z]=n[M_0,z=3]\, .
\label{eqApp|NumberConser}
\end{equation}
The result is shown with a long dashed line in the left panel of Figure~\ref{fig|MergersAppendix},
which shows that such halos grow, on average, by about a factor of $\sim 6$ within
the redshift range $3\le z \le 6$.

We then compute the expected growth
of such halos due to halo and subhalo mergers. To such purpose we estimate
the average number of mergers per unit redshift $dN/dz$ experienced by a given halo of mass $M(z)$ at redshift
$z$ by integrating the halo merger rates
\begin{equation}
\frac{dN}{dz}[M(z),z]=\int_{\xi_{\rm min}}^{\xi_{\rm max}}\frac{B[M(z),\xi,z]}{n[M(z),z]}d\xi \, ,
\label{eqApp|FakMa}
\end{equation}
with $\xi_{\rm min}=0.1$ and $\xi_{\rm max}=1$. We take the halo merger rates per unit
halo $B[M(z),\xi,z]/n[M(z),z]$
from Fakhouri \& Ma (2008), given as empirical fits to the Millennium Simulation, with $n[M(z),z]$
the number density of halos of mass $M(z)$ at redshift $z$. The quantity $B[M(z),\xi,z]$ is the instantaneous merger rate at redshift $z$ of halos with mass
$M(z)$ in units of $\,\, h^4 {\rm \, \, Mpc^{-3}\, dz^{-1}\, d\xi^{-1}}\, M_{\odot}^{-1}$, with $\xi=M_1/M_2$ the mass ratio between the masses of the progenitor merging halos with total mass $M(z)=M_1+M_2$.
The total mass accreted $\Delta M(z)$ on the halo of mass $M(z)$ at each timestep $\Delta z$ during the evolution is then
\begin{equation}
\Delta M(z)=\Delta z \int_{\xi_{\rm min}}^{\xi_{\rm max}}\frac{dN}{dz}[M(z),\xi,z] \times \left[\frac{\xi}{1+\xi} M(z)\right] d\xi\,
\label{eqApp|HaloGrowth}
\end{equation}
with $\xi/(1+\xi) M(z)$ the mass of the (smaller) merging halo.
The solid line in the left panel of Figure~\ref{fig|MergersAppendix} is the total
mass accreted in mergers. As clear from
the bottom panel, which shows the ratio between the cumulative mass grown by
mergers and the total one derived from equation~(\ref{eqApp|NumberConser}), mergers with $0.1<\xi<1$ can account for most ($\sim 65\%$) of the average growth of halos.
\footnote{We note here that the exact value of the adopted $\xi_{\rm min}$
 does not alter our overall conclusions. For example,
 setting $\xi_{\rm min}=0.01$, increases the growth of the parent halo via mergers to $90-95\%$, but it has a negligible impact on the mass growth of the central black hole.}

It is then natural to ask how much
of the central black hole mass is actually contributed by
mergers.
The long-dashed line in the upper right panel of Figure~\ref{fig|MergersAppendix} shows
the total growth of the central black hole derived by assuming
that at each $z$ the black hole has, on average, a mass as given
by equation~(\ref{eq|MbhMhalo}).
In order to estimate the contribution of black hole mergers we, however, cannot naively use equation~(\ref{eqApp|HaloGrowth}). We need to in fact take into
account that when the smaller halo enters the virial radius
of the larger halo, it takes about a dynamical friction
timescale to sink to the center allowing for the central galaxies
(and their black holes) to actually merge.

We therefore follow Shen (2009), and compute the central
galaxy merger rates as
$B_{\rm gal}[M(z),\xi,z]=B[M(z),\xi,z_e]\frac{dz_e}{dz}$, being
$z$ the redshift of the actual merger with the central galaxy and $z_e$ the redshift at which the smaller halo first entered the virial radius of the larger halo. We use the Shen (2009) analytical fit to the function $z_e(z,\xi)$, which reproduces the Jiang et al. (2008) merger timescales well in the range $0.1\le \xi \le 1$. Adopting the results by Jiang et al. (2008) is particularly meaningful for this paper. Their subhalo merger timescales were in fact derived for a suite of high-resolution numerical simulations performed on halos with masses $M>5\times 10^{12}\, h^{-1}$\msun, in the range of interest for our paper, and with a virial mass definition equal to that used in this paper (see \S~\ref{sec|method}). The study by
Boylan-Kolchin et al. (2008), for example, yields somewhat different merging timescales with respect to those found by Jiang et al. (2008; see also Shen 2009). However,
their results were based on parent halos about an order of magnitude less massive
than the ones of interest here, and with a significantly different definition
of virial mass.

The solid line in the upper right panel of Figure~\ref{fig|MergersAppendix} shows the total mass accreted onto the central black hole via mergers. It is clear that subhalo mergers are not the dominant source of growth for massive black holes at high
redshift, as also independently found by several other groups (e.g., Volonteri et al. 2005). The cumulative fraction of black hole mass at $z=3$ grown via mergers is only $\sim 7\%$, reducing to just a few percent at $4<z<4.5$ where most of the tightest constraints from clustering come from (see \S~\ref{sec|results}). Adopting
the Boylan-Kolchin et al. (2008) merging timescales would increase the contribution from mergers by just a factor of $\lesssim 2$.
Note, also, that our estimate is actually an upper limit. In fact, this calculation assumes that all the incoming dark matter substructures actually contain a black hole as massive as what predicted from equation~(\ref{eq|MbhMhalo}) that efficiently merges with the central black hole. Moreover, we have not considered that the delay time for black hole mergers is even longer than those of galaxies (see, e.g., Merritt \& Milosavljevic 2005 for a review), a correction which would further drop the contribution of black hole mergers.

In the lower right panel of Figure~\ref{fig|MergersAppendix} we also show
the predictions of the same model when no delay is considered between halos and central
black hole mergers. In this model, satellite black holes instantly merge
with the central black hole of the parent halo just after the merging
of their host halos. Although this assumption is obviously too simplistic, as it neglects
any dynamical friction time delay, it can be safely regarded
as a secure \emph{upper} limit to the contribution of mergers to
the overall black hole growth. As shown in the right panel
of Figure~\ref{fig|MergersAppendix}, in this extreme model the growth of black hole
mass via mergers is comparable to that via accretion, accounting for about 50-60\%
of the final black hole at $z\sim 3$.

From the study undertaken in this section, we conclude that,
in the physically
plausible case that
a significant dynamical friction time-delay is present between
host halo and central black hole mergers,
it is a good approximation to neglect black hole growth via mergers
in the continuity equation model discussed in \S~\ref{sec|method}.
However, the same is not true for quasar activations. The model
discussed in \S~\ref{sec|method} holds in reproducing both the
quasar luminosity function and quasar clustering
only in the hypothesis that black hole growth via accretion
parallels that of the host dark matter halo with no
time delay between the two. The latter assumption was also
adopted by several other groups
to boost and thus facilitate the match to the high-$z$,
luminous quasar number counts (e.g., Wyithe \& Loeb 2003; Scannapieco \& Ho 2004; Lapi et al. 2006; Shen 2009).

For completeness, however, we present in the next section
the results of a fully self-consistent model that evolves the black hole mass function
through a continuity equation with accretion and mergers, with \emph{no} delay in any of its components.
We will discuss how the outcome of such
a model strengthens our general conclusions.

\section{Models with accretion plus mergers}

Inserting mergers in the continuity equation of equation~(\ref{eq|continEq})
implies a format of the type
\begin{equation}
\frac{\partial n_{\rm BH}(M_{\rm BH},t)}{\partial
t}=-\frac{1}{t_{\rm ef} \ln(10)^2 M_{\rm BH}}\frac{\partial
\Phi(L,z)}{\partial \log L} + S_{\rm in} - S_{\rm out}\,
\label{eqApp|contEqMergers}
\end{equation}
where
\begin{eqnarray}
S_{\rm in}=\frac{1}{2}\, \int_{\xi_{\rm min}}^{1} d\xi \left( \frac{P_{\rm merg,z}}{\Delta t}(\xi, M)n_h \left[M \left(M_{\rm BH}'=\frac{\xi M_{\rm BH}}{1+\xi}, z \right),z \right]\frac{dM}{dM_{\rm BH}'}\frac{dM_{\rm BH}'}{dM_{\rm BH}} \right) + \\ \nonumber
d\xi \left(\frac{P_{\rm merg,z}}{\Delta t}(\xi, M)n_h \left[M \left(M_{\rm BH}''=\frac{M_{\rm BH}}{1+\xi}, z \right),z \right]\frac{dM}{dM_{\rm BH}''}\frac{dM_{\rm BH}''}{dM_{\rm BH}}  \right)\, ,
\label{eqApp|Sin}
\end{eqnarray}
and
\begin{eqnarray}
S_{\rm out}=\frac{1}{2}\, \int_{\xi_{\rm min}}^{1} d\xi \left( \frac{P_{\rm merg,z}}{\Delta t}(\xi, M)n_h \left [M \left(M_{\rm BH}'=\frac{1+\xi}{\xi}M_{\rm BH}, z \right),z \right]\frac{dM}{dM_{\rm BH}'}\frac{dM_{\rm BH}'}{dM_{\rm BH}} \right) + \\ \nonumber d\xi \left(\frac{P_{\rm merg,z}}{\Delta t}(\xi, M)n_h \left[M \left(M_{\rm BH}''=(1+\xi)M_{\rm BH}, z \right),z \right]\frac{dM}{dM_{\rm BH}''}\frac{dM_{\rm BH}''}{dM_{\rm BH}}  \right)\, ,
\label{eqApp|Sout}
\end{eqnarray}
are, respectively, the merger rate of incoming smaller mass black holes with
 mass $M_{\rm BH}'=M_{\rm BH}\xi/(1+\xi)$ and $M_{\rm BH}''=M_{\rm BH}/(1+\xi)$ that
 merge into a black hole of final mass $M_{\rm BH}$, and the merger rate of black holes with initial mass $M_{\rm BH}$ that merge into more massive black holes of mass
 $M_{\rm BH}'=M_{\rm BH}(1+\xi)/\xi$ and $M_{\rm BH}''=M_{\rm BH}(1+\xi)$ (in both Eqs.~[C2] and [C3] we set $\xi_{\rm min}=0.1$, and add the factor
of $1/2$ to avoid double counting).

If we assume that no delay is present between
the mergers of the black holes and their parent halos, the probability of black hole
mergers per unit time is simply given by the
halo merger rate adopted above, i.e.,
\begin{equation}
\frac{P_{\rm merg,z}}{\Delta t}(\xi, M)n_h [M (M_{\rm BH}, z),z]=B_h[M,\xi(z),z]\, .
\label{eqApp|ratemergers}
\end{equation}
By simply knowing, at each timestep, the mapping between infalling halo mass and its central black hole (given by Eq.~[\ref{eq|MbhMhalo}]), we can then compute the expected average rate
for any black hole merger event.
By further assuming the AGN luminosity function to be known from observations (we here adopt the analytical derivation by SWM),
we can simply integrate the right-hand side of equation~(\ref{eqApp|contEqMergers}) and derive
the black hole function at all redshifts.

\begin{figure}
\epsscale{1.0} \plottwo{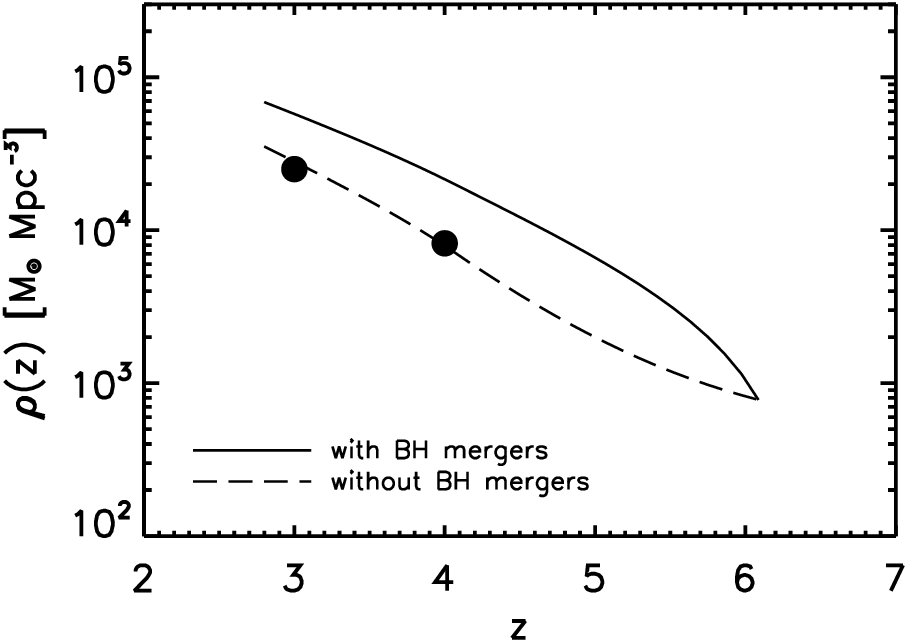}{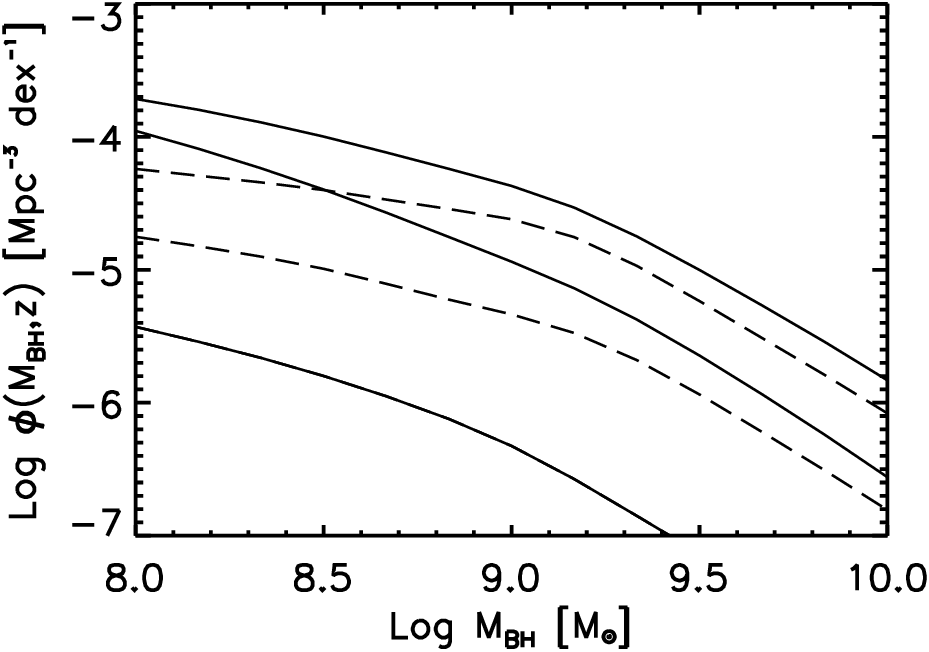} \caption{\emph{Left}: accreted black hole mass density above $\log M_{\rm BH}/M_{\odot} > 8$ as a function of redshift, for the reference model characterized by $\lambda=0.25$ and $\epsilon=0.15$, with no mergers (\emph{long-dashed} line), and with mergers (\emph{solid} line); the solid, filled circles are the $z=3$ and $z=4$ black hole mass density obtained via Eq.~[\ref{eq|rhoBHatgivenz}] by assuming a duty cycle
of $P_0=0.25$ and $P_0=0.5$, respectively, as in the reference model (Figs.~\ref{fig|dutycycle} and \ref{fig|r0zDuty}).  \emph{Right}: comparison between the resulting black hole mass functions for the no-merger (\emph{long-dashed}) and merger (\emph{solid}) models, at three different redshifts, from bottom to top, $z=6, 4, 3$.}
\label{fig|ContinuityWithMergers}
\end{figure}

The result is shown in Figure~\ref{fig|ContinuityWithMergers}, the left panel of which shows the integrated black hole mass density $\rho(M_{\rm BH}>10^8 M_{\odot},z)$, in the mass and redshift range of interest here, for the reference model ($\lambda =0.25$, $\alpha=1$, $\epsilon=0.15$) with
no mergers (long-dashed lines), and with mergers (solid lines), as labeled, while the right panel
compares the resulting black hole mass functions for the two models at three different redshifts, from bottom to top, $z=6, 4, 3$. The two filled circles in the left panel mark the expected black hole mass
density at $z=3$ and $z=4$ expected from equation~(\ref{eq|rhoBHatgivenz}), adopting a duty cycle
of $P_0=0.25$ and $P_0=0.5$, respectively, for a model consistent with the measured quasar clustering (see Figure~\ref{fig|r0zDuty}). As above, we here neglect any source of scatter in the relation between luminosity and halo mass to maximize the predicted clustering for a given model.
We find that, irrespective of the differences in the AGN luminosity function adopted here and in the main text, the results for the pure accretion model match those presented in Figures~\ref{fig|r0zDuty}. When mergers are included, the estimated black hole mass density above $10^8 M_{\odot}$ is larger by a factor of $\gtrsim 2$, than the one from accretion alone.
We should emphasize that we consider this factor of two to be an
extreme upper limit on the impact of mergers, since it ignores even the
effects of dynamical friction delay, which we have shown in the
previous Appendix to drastically reduce the black hole merger growth.
More importantly, however, any growth of the black hole mass function
via mergers {\it exacerbates} the tension we have highlighted
between high-redshift quasar clustering and the luminosity function.
Mergers add mass to the black hole population without associated luminosity,
so reproducing the observed luminosity function with the black
hole population implied by the black hole-halo relation requires a still
higher radiative efficiency.

%

\end{appendix}


\end{document}